\documentclass[12pt]{article}
\usepackage{amssymb}

\usepackage{amsmath}
\usepackage{latexsym}
\usepackage{epsfig}

\newif{\ifcomentarios}
\comentariosfalse

\newtheorem{theorem}{Theorem}
\newtheorem{lemma}[theorem]{Lemma}

\newtheorem{proposition}[theorem]{Proposition}

\newtheorem{criterium}[theorem]{Criterium}

\newtheorem{remark}[theorem]{Remark}

\begin{document}

\author{\textbf{Leonardo F. Guidi}\thanks{%
Supported by FAPESP under grant $\#98/10745-1$. E-mail: \textit{%
guidi@if.usp.br}. \ \ \ \ \ \ } \quad \&\quad \textbf{Domingos H. U.
Marchetti} \thanks{%
Partially supported by CNPq, FINEP and FAPESP. E-mail: \textit{%
marchett@if.usp.br}} \\
Instituto de F\'{\i}sica\\
Universidade de S\~{a}o Paulo \\
Caixa Postal 66318\\
05315 S\~{a}o Paulo, SP, Brasil}
\title{Renormalization Group Flow \\
of the Two-Dimensional Hierarchical Coulomb Gas}
\date{}
\maketitle

\begin{abstract}
We consider a quasilinear parabolic differential equation associated with
the renormalization group transformation of the two--dimensional
hierarchical Coulomb system in the limit as the size of the block $%
L\downarrow 1$. We show that the initial value problem is well defined in a
suitable function space and the solution converges, as $t\rightarrow \infty$,
to one of the countably infinite equilibrium solutions. The $j$--th 
nontrivial equilibrium solution bifurcates from the trivial one at $\beta
_{j}=8\pi /j^{2}$, $j=1,2,\ldots $. These solutions are fully described and
we provide a complete analysis of their local and global stability for all
values of inverse temperature $\beta >0$. Gallavotti and Nicol\'{o}'s
conjecture on infinite sequence of ``phases transitions'' is also addressed.
Our results rule out an intermediate phase between the plasma and the
Kosterlitz--Thouless phases, at least in the hierarchical model we consider.
\end{abstract}


\section{Introduction}

\setcounter{equation}{0} \setcounter{theorem}{0}

We consider, for each $\beta >0$, the partial differential equation

\begin{equation}
u_{t}-\frac{\beta }{4\pi }(u_{xx}-u_{x}^{2})-2u=0\,  \label{pde}
\end{equation}
on $\mathbb{R}_{+}\times \left( -\pi ,\pi \right) $ with periodic boundary
condition, $u(t,-\pi )=u(t,\pi )$ and $u_{x}(t,-\pi )=u_{x}(t,\pi )$, in the
space of even functions, satisfying an additional condition
$u(t,0)=0$\footnote{This is assured by a Lagrange multiplier (see Remark 
\ref{Lagrange}).}. We 
show that the initial value problem is well defined in an appropriate
function space $\mathcal{B}$ and the solution exists and is unique for all $
t>0$. Furthermore, as $t\rightarrow \infty $, the solution converges in $
\mathcal{B} $ to one of the (equilibrium) solutions $\phi $ of 
\begin{equation}
\frac{\beta }{4\pi }\left( \phi ^{\prime \prime }-(\phi ^{\prime
})^{2}\right) +2\phi =0\,,  \label{stateq}
\end{equation}
with $\phi (-\pi )=\phi (\pi )$ and $\phi ^{\prime }(-\pi )=\phi ^{\prime
}(\pi )$. For $\beta >8\pi $, $\phi _{0}\equiv 0$ is the (globally)
asymptotically stable solution of (\ref{pde}). For $\beta <8\pi $ such that $
8\pi /\left( k+1\right) ^{2}\leq \beta <8\pi /k^{2}$ holds for some $k\in 
\mathbb{N}_{+}$, $\phi _{0}$ is unstable and there exist $2k$ non--trivial
equilibria solutions $\phi _{1}^{\pm },\ldots ,\phi _{k}^{\pm }$ of
(\ref{stateq}) among which $\phi _{1}^{\pm }$ are the only asymptotically
stable ones.  

The aim of the present work is to show that, for $j\geq 1$, $\phi _{j}^{\pm
} $ have a $\left( j-1\right) $--dimensional unstable manifold $\mathcal{M}
_{j}\subset \mathcal{B}$ so $\phi _{j}^{\pm }$ are more stable than $\phi
_{j^{\prime }}^{\pm }$ if $j<j^{\prime }$. As a consequence, there exists a
dense open set of initial conditions in $\mathcal{B}$ such that $\phi
_{1}^{+}$ ($\phi _{1}^{-}$ is not physically admissible) is the non--trivial
stable solution for all $\beta <8\pi $.

Our description of equation (\ref{pde}) is motivated by two distinct goals.
Firstly, it provides a new example of nonlinear parabolic differential
equation by which a geometric theory can be carried out (see e.g. Henry \cite
{H}). According to this theory, the above scenario can be stated as follows:
there exist a sufficient large ball $\mathcal{B}_{0}\subset \mathcal{B}$
about the origin such that, if $u(t,\mathcal{B}_{0})$ denotes the set of
points reached at time $t$ starting from any initial function in $\mathcal{B}
_{0}$, then the invariant set $\bigcap_{t\geq 0}u(t,\mathcal{B}_{0})$
coincides with the $k$--dimensional unstable manifold $\mathcal{K}%
_{k}=\bigcup_{0\leq j\leq k}\mathcal{M}_{j}=\overline{\mathcal{M}_{0}}$
provided $8\pi /(k+1)^{2}\leq \beta <8\pi /k^{2}$.

Secondly, the solution of the initial value problem (\ref{pde}) describes
the renormalization group (RG) flow of the effective potential in the
two--dimensional hierarchical Coulomb system and the stationary solutions $
\left\{ \phi _{j}^{+}\right\} $, the fixed points of RG, contain
informations on its critical phenomena.

The analysis of equation (\ref{pde}) presented here can hopefully bring some
light to a question raised by Gallavotti and Nicol\'{o} \cite{GN} on the
``screening phase transitions'' in two--dimensional Coulomb systems. The
existence of infinitely many thresholds of ``instabilities'' found in the
Mayer series at inverse temperature $\beta _{n}=8\pi (1-1/(2n))$, $n\in 
\mathbb{N}_{+}$, indicates, according to the authors, a sequence of
``intermediate'' phase transitions from the plasma phase ($\beta \leq \beta
_{1}=4\pi )$ to the multipole phase ($\beta \geq \beta _{\infty }=8\pi $).
They conjectured that some partial screening takes place when the inverse
temperature decreases from $8\pi $ to $4\pi $, which prevents the formation
of neutral multipole of order larger than $2n$ where $n$ is the integer part
of $1/(2-\beta /4\pi )$ (dipoles are the last to be prevented at $4\pi $).

The Kosterlitz--Thouless phase (multipole phase) was established by
Fr\"{o}hlich--Spencer \cite{FS} and extended up to $8\pi $ by one of the
present authors and A. Klein \cite{MK}. Debye screening (plasma phase) was
only proved for sufficiently small $\beta <<4\pi $ \cite{BF}. Study of the
region $[4\pi ,8\pi ]$ began with the work by Benfatto, Gallavotti and
Nicol\'{o} \cite{BGN} on the ultraviolet collapses of neutral clusters in
the Yukawa gas which served as a base for the results in \cite{GN}. It seems
improbable, on the light of the present knowledge, that a conclusive answer
to the Gallavotti--Nicol\'{o} conjecture will come up soon. It may be noted,
however, that the scenario of an intermediate phase, which has challenged
the conventional picture due to Jose \emph{et al} \cite{JKKN}, has been
contested by Fisher \emph{et al } \cite{FLL} based on
Debye--H\"{u}ckel--Bjerrum theory and by Dimock and Hurd \cite{DH} who have
reinterpreted the ultraviolet collapses in the Yukawa gas.

The Kosterlitz--Thouless phase is manifested in the hierarchical model as a
bifurcation from the trivial solution \cite{MP}. Our results rule out the
existence of further phase transitions since no other bifurcation arises
from the stable solution (see Theorem \ref{lstb} on the stability of $\phi
_{1}^{+}$).

Even though the existence of the invariant unstable manifold $\mathcal{K}
_{k} $ may provide a suitable explanation to the appearance of
Gallavotti--Nicol\'{o}'s thresholds, the nature (and location) of the
instabilities in the hierarchical Coulomb gas differs substantially from the
one we have just described, because neutral multipoles cannot be formed in
the hierarchical model. We believe, however, our investigation may be
helpful for the plasma phase. Numerical analysis shows the stable solution $
\phi _{1}^{+}$ looks like the Debye--H\"{u}ckel potential $\phi _{DH}=(2\pi
/\beta )\,x^{2}$ in $(-\pi ,\pi )$ right after the transition takes place
(see Remark \ref{sinegordon}).

As in \cite{F}, the renormalization group (GR) flow (\ref{pde}) may be
derived from the block--spin RG transformation of a two--dimensional
hierarchical Coulomb system in the limit as the block size $L\downarrow 1$.
This procedure, called \emph{local potential approximation}, has been
discussed by Felder \cite{F} in the context of Dyson's hierarchical model,
whose partial differential equation, 
\begin{equation}
u_{t}-\frac{1}{2}u_{xx}+\frac{d-2}{2}x\,u_{x}-d\,u+\frac{1}{2}u_{x}^{2}=0\,,
\label{pde-f}
\end{equation}
coincides with (\ref{pde}) when his dimensional parameter $d=2$ if $\beta $
is equal to $2\pi $ (without boundary conditions). Felder showed that
(\ref{pde-f}) has global stationary solutions $u_{2n}^{\ast }$ on $\mathbb{R}$ 
for $2<d<d_{n}$ with $u_{2n}^{\ast }(x)\rightarrow 0$ as $d\uparrow d_{n}$
and calculated their profile. Here, $d_{n}=2+2/(n-1)$, $n=2,3,\ldots $, is
the sequence of thresholds where nontrivial fixed points are expected to
appear as a bifurcation from the trivial solution. We mean by global
solution one which doesn't blow up at finite $x$.

The present paper begins with a derivation of equation (\ref{pde}) in
Section \ref{rgflow}. The existence, uniqueness and continuous dependence on
the initial value are presented in Section \ref{EU} and the precise
statements are given in Theorems \ref{thivp} and \ref{cdi}. We describe all
global solutions of (\ref{stateq}) completely in Section \ref{SS}. Due to
smoothness and the periodic condition, blow--up of an admissible stationary
solution is impossible. We show that the non--trivial stationary solution
for $\beta <8\pi $ is \emph{unique} modulo solutions with period $2\pi /j$, $
j=2,3,\ldots $, which are responsible for the existence of the unstable
manifold (see Theorem \ref{existence}). Finally, we analyze in detail the
local and global stability of equilibrium solutions of (\ref{pde}) in
Section \ref{Stability}. The main results are stated in Theorems \ref{lstb}
and \ref{gstb}.


\section{The Flow Equation \label{rgflow}}

\setcounter{equation}{0} \setcounter{theorem}{0}

This section is devoted to the derivation of (\ref{pde}) from the RG
transformation of two--dimensional hierarchical Coulomb system. We begin
with a brief review of this model.

A Coulomb system is an ensemble of two species (for simplicity) of charged
particles, interacting via a two--body Coulomb potential $V$. In the grand
canonical ensemble the total number of particles fluctuates around a mean
value determined by the particle activity $z$. It will become clear that the
charge ensemble, rather than the particle ensemble, is more appropriate for
RG transformation.

A configuration $q$ of this system is a function $q:\Lambda \subset \mathbb{Z
}^{2}\longrightarrow \mathbb{Z}$ which associates to each site $x$ of the
lattice $\Lambda $ the total charge $q(x)$ at this position.

To each configuration we introduce two functionals: the total energy $E: 
\mathbb{Z}^{\Lambda }\longrightarrow \mathbb{R}_{+}$, 
\begin{equation}
E(q)=\frac{1}{2}\sum_{x,y\in \Lambda }q(x)\,V(x,y)\,q(y)  \label{E}
\end{equation}
(self--energy is included) and an ``a priori'' weight $F:\mathbb{Z}^{\Lambda
}\longrightarrow \mathbb{R}_{+}$, 
\begin{equation}
F(q)=\prod_{x\in \Lambda }\lambda (q(x))  \label{F}
\end{equation}
defined for positive real valued functions $\lambda $.

The equilibrium Gibbs measure $\mu _{\Lambda }:\mathbb{Z}^{\Lambda
}\longrightarrow \mathbb{R}_{+}$ is thus given by 
\begin{equation}
\mu _{\Lambda }(q):=\frac{1}{\Xi _{\Lambda }}F(q)\,e^{-\beta \,E(q)}
\label{mu}
\end{equation}
where $\beta $ is the inverse temperature and 
\begin{equation}
\Xi _{\Lambda }=\sum_{q\in \mathbb{Z}^{\Lambda }}F(q)\,e^{-\beta \,E(q)}
\label{Xi}
\end{equation}
is the grand partition function.

It has been shown (see e.g. \cite{FS}) that the standard Coulomb system in
the grand canonical ensemble with particle activity $z$ has charge activity
given by $\lambda (q)=I_{q}(2z)$, where $I_{q}$ is the $q$--th modified
Bessel function. If $\lambda (q)=\delta _{q,0}+z\left( \delta _{q,1}+\delta
_{q,-1}\right) $, $\Xi _{\Lambda }$ is the grand canonical ensemble of
charged particles with hard core.

Let us introduce our hierarchical model as proposed in ref. \cite{MP}. The
potential $V$ in (\ref{E}) is replaced by a function 
\begin{equation*}
V_{h}(x,y)=-\frac{1}{2\pi }\ln d_{h}(x,y)\,,
\end{equation*}
given by the asymptotic behavior of \ the two--dimensional Coulomb potential
with the Euclidean distance $\left| x-y\right| $ replaced by hierarchical
distance 
\begin{equation}
d_{h}(x,y):=L^{N(x,y)}\,,  \label{dh}
\end{equation}
defined for an integer $L>1$, where 
\begin{equation}
N(x,y):=\inf \left\{ N\in \mathbb{N}_{+}:\left[ \frac{x}{L^{N}}\right] =
\left[ \frac{y}{L^{N}}\right] \right\}  \label{N}
\end{equation}
and $[z]\in \mathbb{Z}^{2}$ has components the integer part of the
components of $z\in \mathbb{R}^{2}$. Notice that $d_{h}$ is not invariant by
translations.

Now, given an integer number $N>1$ , let $\Lambda =\Lambda
_{N}=[-L^{N},L^{N}-L^{N-1}]^{2}\cap \mathbb{Z}^{2}$ and define, for each
configuration $q\in \mathbb{Z}^{\Lambda }$, the block configuration $%
q^{1}:\Lambda _{N-1}\longrightarrow \mathbb{Z}$, 
\begin{equation}
q^{1}(x)=\sum_{\underset{i=1,2}{0\leq y_{i}<L}}q(Lx+y)\,.  \label{q1}
\end{equation}

The renormalization group transformation $\mathcal{R}$ acting on the space
of Gibbs measures (\ref{mu}), 
\begin{equation}
\mu _{\Lambda _{N-1}}^{1}(q^{1})=[\mathcal{R}\mu _{\Lambda
_{N}}](q^{1}):=\sum_{\underset{q^{1}\mathrm{fixed}}{q\in \mathbb{Z}^{\Lambda
_{N}}:}}\mu _{\Lambda _{N}}(q)\,,  \label{mu1}
\end{equation}
involves an integration over the fluctuations about $q^{1}$ following by a
rescaling back to the original lattice.

As it has been shown in \cite{MP}, the RG transformation $\mathcal{R}$
preserves the form of the Gibbs measure in the grand canonical ensemble of
charges. The measure $\mu _{\Lambda _{N-1}}^{1}$ is thus given by (\ref{mu})
with the ``a priori weight'' $F$ replaced by 
\begin{equation}
F^{1}(q^{1})=\prod_{x\in \Lambda _{N-1}}\lambda ^{1}(q^{1}(x))  \label{F1}
\end{equation}
where 
\begin{equation}
\lambda ^{1}(p)=L^{-\beta p^{2}/(4\pi )}\underset{L^{2}-\mathrm{times}}{(%
\underbrace{\lambda \star \lambda \star \cdots \star \lambda })}(p)
\label{lambda}
\end{equation}
with $(\lambda \star \varrho )(p)=\sum\limits_{q\in \mathbb{Z}}\lambda
(p-q)\,\varrho (q)$. Note that $\Xi _{\Lambda _{N}}(\lambda )=\Xi _{\Lambda
_{N-1}}(\lambda ^{1})$.

\begin{remark}
A peculiar feature of hierarchical models is the reduction of the measure
space where $\mathcal{R}$ acts to local functions. The RG transformation (%
\ref{mu1}) induces a transformation $\lambda ^{1}=r\lambda $ given by (\ref
{lambda}) on the space of infinite sequences. Note that the space $\ell _{1}(%
\mathbb{Z})$ of summable sequences is closed by the $r$ transformation: $%
\,(\lambda \star \lambda )\in \ell _{1}(\mathbb{Z})$ if $\lambda \in \ell
_{1}(\mathbb{Z})$ by the Hausdorff-Young inequality.
\end{remark}

In order to take $L\downarrow 1$ limit of the RG transformation $r$ it is
convenient to write the system in the \textit{sine--Gordon representation.}
Fourier transforming (\ref{lambda}), 
\begin{equation*}
\widehat{\lambda }(\varphi )=\sum_{q\in \mathbb{Z}}\,\lambda
(q)\,e^{iq\varphi }\,,
\end{equation*}
and using the convolution theorem, yields 
\begin{equation}
\widehat{\lambda ^{1}}(\varphi )=\widehat{r\lambda }(\varphi )=\frac{1}{2\pi 
}\int_{-\pi }^{\pi }\vartheta (\varphi -\tau )\,\,\widehat{\lambda }%
^{L^{2}}(\tau )\,d\tau  \label{l-tilde}
\end{equation}
where 
\begin{eqnarray}
\vartheta (\varphi ) &=&\sum_{q\in \mathbb{Z}}\,L^{-\beta q^{2}/(4\pi
)}\,e^{iq\varphi }  \notag \\
&=&\frac{1}{(\beta \ln L)^{1/2}}\sum_{n\in \mathbb{Z}}e^{-\pi (\varphi +2\pi
n)^{2}/(\beta \ln L)}  \label{nu}
\end{eqnarray}
by the Poisson formula.

Plugging (\ref{nu}) into (\ref{l-tilde}) and changing the variable $\zeta
=\tau +2\pi n$, equation (\ref{l-tilde}) can be written as 
\begin{equation}
\widehat{r\lambda }(\varphi )=\left( \nu \ast \widehat{\lambda }%
^{L^{2}}\right) (\varphi )  \label{l-t1}
\end{equation}
where $\nu \ast $ means convolution by a Gaussian measure with mean zero and
variance $\beta \ln L/(2\pi )$: 
\begin{eqnarray}
(\nu \ast f)(\varphi ) &=&\left( \beta \ln L\right) ^{-1/2}\int_{-\infty
}^{\infty }d\zeta \,\,e^{-\pi \left( \varphi -\zeta \right) ^{2}/(\beta \ln
L)}\,f(\zeta )  \label{gauss} \\
&=&e^{\left( \beta \ln L/4\pi \right) \left( d^{2}/d\varphi ^{2}\right)
}\,f(\varphi )\,\,,  \notag
\end{eqnarray}
where in the second form of the Gaussian convolution we have used Wick's
theorem.

Note that (\ref{l-t1}) is precisely the RG transformation derived by
Gallavotti who has started directly from the sine-Gordon representation.

In order to let the block size $L$ to $1$, we introduce a variable $%
t:=n\,\ln L$ which keeps track of the number of times the RG transformation (%
\ref{mu1}) has to be iterated in order to bring two sites at hierarchical
distance $L^{n}$ to $\mathcal{O}(1)$ distance. We shall take the limit $%
L\downarrow 1$ together with $n\rightarrow \infty $ maintaining $t$ fixed.

Define 
\begin{equation}
u(t,x)=-\ln \widehat{\lambda ^{n}}(x)  \label{utx}
\end{equation}
where $\widehat{\lambda ^{n}}=\widehat{r^{n}\lambda }$ denotes the $n$ --th
iteration of the transformation (\ref{l-t1}). If one writes $t^{\prime
}=(n+1)\ln L$ then, by taking the logarithm and using (\ref{utx}), equation
(\ref{l-t1}) reads  
\begin{eqnarray}
u(t^{\prime },x) &=&-\ln \left\{ \exp \left( \frac{\beta t}{4\pi n}\frac{%
d^{2}}{dx^{2}}\right) \exp \left( -e^{2t/n}u(t,x)\right) \right\}  \notag \\
&=&u(t,x)-\ln \left\{ 1+\frac{t}{n}\left( \frac{\beta }{4\pi }\left(
u_{x}^{2}(t,x)-u_{xx}(t,x)\right) -2u(t,x)\right) +\mathcal{O}\left( \frac{1
}{n^{2}}\right) \right\}  \label{uutx} \\
&=&u(t,x)+\frac{t}{n}\left( \frac{\beta }{4\pi }\left(
u_{xx}(t,x)-u_{x}^{2}(t,x)\right) +2u(t,x)\right) +\mathcal{O}\left( \frac{1
}{n^{2}}\right)  \notag
\end{eqnarray}
which, combined with 
\begin{eqnarray}
u_{t}(t,x) &=&\lim_{t^{\prime }\downarrow t}\frac{u(t^{\prime },x)-u(t,x)}{
t^{\prime }-t}  \notag \\
&=&\lim_{n\rightarrow \infty }\frac{n}{t}\left( u(t^{\prime
},x)-u(t,x)\right) \,,  \label{utxutx}
\end{eqnarray}
yields equation (\ref{pde}).


\section{Existence, Uniqueness and Continuous Dependence \label{EU}}

\setcounter{equation}{0} \setcounter{theorem}{0}

In this section the existence, uniqueness and continuous dependence on the
initial value of equation (\ref{pde}) will be established by Picard's
theorem for Banach spaces.

To avoid the appearance of zero modes upon linearization, we differentiate
(\ref{pde}) with respect to $x$ and consider the equation for $v=u_{x}$,  
\begin{equation}
v_{t}-\frac{\beta }{4\pi }\left( v_{xx}-2v\,v_{x}\right) -2v=0\,,
\label{dpde}
\end{equation}
with $v\left( t,-\pi \right) =v\left( t,\pi \right) $ and $v_{x}\left(
t,-\pi \right) =v_{x}\left( t,\pi \right) $, in the subspace of odd
functions and initial value $v(0,\cdot )=v_{0}$. Note that the operator
defined by the l. h. s. of (\ref{dpde}) preserves this subspace.

Before we proceed, we have the following

\begin{remark}
\label{Lagrange}The ``a priori weight'' $\lambda (t,q):=\lambda ^{n}(q)\,$\
at scale $t=n\ln L$, is a positive \textbf{symmetric}, $\lambda
(t,q)=\lambda (t,-q)$, sequence of real numbers and has to be normalized at
all scales. In \cite{MP} equation (\ref{lambda}) was redefined so that $%
\lambda ^{n}(0)=1$ holds for all $n$. Here, the appropriated normalization\
is given by 
\begin{equation*}
\sum_{q\in \mathbb{Z}}\lambda (t,q)=1\,,
\end{equation*}
since, in view of equation (\ref{utx}), this leads to the condition $
\widetilde{u}\left( t,0\right) =0$, which is already imposed for all $t$ if 
\begin{equation}
\widetilde{u}(t,x)=\int_{0}^{x}v(t,y)\,dy  \label{utilde}
\end{equation}
with $v(s,x)$ an odd solution of (\ref{dpde}). From (\ref{utilde}), we have 
\begin{eqnarray}
\widetilde{u}_{t} &=&\int_{0}^{x}v_{t}(t,y)\,dy  \notag \\
&=&\int_{0}^{x}\left[ \alpha \left( \widetilde{u}_{xx}-\widetilde{u}
_{x}^{2}\right) +2\widetilde{u}\right] _{x}\,dy  \notag \\
&=&\alpha \left( \widetilde{u}_{xx}-\widetilde{u}_{x}^{2}\right) +2
\widetilde{u}-\alpha \widetilde{u}_{xx}(t,0)  \label{utildeq}
\end{eqnarray}
where $\widetilde{u}_{x}(t,0)=v(t,0)=0$ by parity. Note that $\widetilde{u}
(t,x)=-\ln \widetilde{\lambda ^{n}}(x)+\ln \widetilde{\lambda ^{n}}(0)$ also
satisfies (\ref{utildeq}) by equations (\ref{uutx}) and (\ref{utxutx}).
Moreover, note that there is a one--to--one correspondence between the
solution of (\ref{pde}) and the solution of (\ref{utildeq}), with the same
initial value $u_{0}$, given by 
\begin{equation}
\widetilde{u}(t,x)=u(t,x)-u(t,0)  \label{utildeu}
\end{equation}
and 
\begin{equation}
u(t,x)=\widetilde{u}(t,x)+\alpha \int_{0}^{t}e^{2(t-s)}\widetilde{u}
_{xx}(s,0)\,ds\,\,,  \label{uutilde}
\end{equation}
where $\alpha \widetilde{u}_{xx}(t,0)$ is the required Lagrange multiplier
introduced in (\ref{utildeq}) to assure that $\widetilde{u}(t,0)=0$ (see
comments after equation (\textsl{1.1}$^{\prime }$) in ref. \cite{F}). This
correspondence will be useful in Section \ref{Stability}.

Because the standard initial condition $u_{0}(x)=z\left( 1-\cos x\right) $
satisfies $\ u_{0}^{\prime }(0)=u_{0}^{\prime }(\pi )=0$, equation (\ref
{dpde}) may equivalently be considered on $\left( 0,\pi \right) $ with
Dirichlet boundary conditions $v\left( t,0\right) =v\left( t,\pi \right) =0$.
\end{remark}

Another reason for considering (\ref{dpde}) instead of (\ref{pde}) is the
fact that the nonlinearity $2v\,v_{x}$ is more suitable than $u_{x}^{2}$ for
the analysis of equilibrium solutions and corresponding stabilities given in
the next sections.

The boundary and initial value problem (\ref{dpde}) may be written as an
ordinary differential equation 
\begin{equation}
\frac{dz}{dt}+Az=F(z)  \label{ode}
\end{equation}
in a conveniently defined Banach space $\mathcal{B}$ where 
\begin{equation}
Az=-\alpha z\,^{\prime \prime }-2z\qquad \mathrm{and}\qquad F(z)=-2\alpha
z^{\prime }z\,,  \label{A}
\end{equation}
with $\alpha =\beta /(4\pi )$ and initial value $z(0)=z_{0}$.

The linear operator $A$ is defined on the space $C_{\mathrm{o,p}}^{2}$ of
smooth odd and periodic real--valued functions in $[-\pi ,\pi ]$,\footnote{
From here on, the subindexes in $C_{\mathrm{o,p}}^{2}$, $L_{\mathrm{o,p}
}^{2} $, $L_{\mathrm{e,p}}^{2}$, $H_{\mathrm{e,p}}^{1}$ and \textit{etc.},
indicate spaces of odd and periodic (o,p) or even and periodic (e,p)
functions.} with inner product $\left( f,g\right) :=\displaystyle\int_{-\pi
}^{\pi }f(x)\,g(x)\,dx$. Because of $\left( f,Ag\right) =\left( Af,g\right)
\,$, $A$ may be extended to a self--adjoint operator in $L_{\mathrm{o,p}
}^{2}\left( -\pi ,\pi \right) $. The domain $D(A)$ of $A$ is 
\begin{equation*}
D(A)=\left\{ f\in L_{\mathrm{o,p}}^{2}\left( -\pi ,\pi \right) :Af\in L_{
\mathrm{o,p}}^{2}\left( -\pi ,\pi \right) \right\}
\end{equation*}
and the spectrum of $A$, 
\begin{equation}
\sigma (A)=\left\{ \lambda _{n}=\alpha n^{2}-2,\,n\in \mathbb{N}_{+}\right\}
\,,  \label{spectrum}
\end{equation}
consists of simple eigenvalues with corresponding eigenfunctions $\phi
_{n}(x)=\left( 1/\pi \right) ^{1/2}\sin \,nx\,$.

Let $A_{1}$ denote a positive definite linear operator given by $A$ if $
\alpha >2$ and $A+aI$ for some $a>2-\alpha $, otherwise. The following
properties also hold for $A$ given by the closure in $L_{\mathrm{o,p}
}^{q}\left( -\pi ,\pi \right) $, $1\leq q<\infty $, of the operator $\left.
\left( -\alpha \,d^{2}/dx^{2}-2\right) \right| _{\mathcal{C}_{\mathrm{o,p}
}^{2}}$.

\begin{enumerate}
\item  The operator $A$ generates an analytic semi--group $T(t)=e^{-tA}$
given by the formula 
\begin{equation*}
T(t)=\frac{1}{2\pi i}\int_{\Gamma }\frac{1}{\lambda +A}\,e^{\lambda
t}\,d\lambda 
\end{equation*}
where $\Gamma $ is a contour in the resolvent set of $A$ with $\arg \lambda
\longrightarrow \pm \theta $, $\pi /2<\theta <\pi $, as $\left| \lambda
\right| \rightarrow \infty $. From this, we have 
\begin{equation}
\left\| e^{-tA}\right\| \leq C\,e^{-ct}\hspace{0.5in}\mathrm{and}\hspace{
0.5in}\left\| Ae^{-tA}\right\| \leq \frac{C}{t}\,e^{-ct}  \label{bounds}
\end{equation}
for $t>0$, $c<\inf_{\lambda }\sigma \left( A\right) $ and $C<\infty $.

\item  Given $\gamma \geq 0$, let the fractional power of $A_{1}$ be given
by 
\begin{equation*}
A_{1}^{-\gamma }=\frac{1}{\Gamma (\gamma )}\int_{0}^{\infty }\,t^{\gamma
-1}\,e^{-A_{1}t}\,dt
\end{equation*}
and define $A_{1}^{\gamma }=\left( A_{1}^{-\gamma }\right) ^{-1}$. $
A_{1}^{-\gamma }$ is a bounded operator (compact if $\gamma >0$) with $
A_{1}^{-1/2}\left( d/dx\right) $ and $\left( d/dx\right) A_{1}^{-1/2}$
bounded in the $L_{\mathrm{o,p}}^{2}\left( -\pi ,\pi \right) $ norm. In
addition, for $\gamma >0$, $A_{1}^{\gamma }$ is closely defined with the
inclusion $D(A_{1}^{\gamma })\subset D(A_{1}^{\tau })$ if $\gamma >\tau $.
\end{enumerate}

It thus follows from $1.$ and $2.$ (see e.g. \cite{H}) 
\begin{equation}
\left\| A_{1}^{\gamma }e^{-tA_{1}}\right\| \leq \frac{C_{\gamma }}{t^{\gamma
}}\,e^{-ct}  \label{b1}
\end{equation}
holds for $0<\gamma <1$, $t>0$. Here $C_{\gamma }$ is bounded in any compact
interval of $\left( 0,1\right) $ and also bounded as $\gamma \searrow 0$.
Note that, if the operator norm is induced by the $L^{2}$--norm, equation
(\ref{b1}) hold with  
\begin{equation}
C_{\gamma }=\sup_{n\in \mathbb{N}_{+}}\left| \left( t\lambda _{n}\right)
^{\gamma }\,e^{-t\lambda _{n}}\right| \leq \sup_{r>tc}\left| r^{\gamma
}\,e^{-r}\right| \leq \left( \frac{\gamma }{e}\right) ^{\gamma }\,,
\label{Cgamma}
\end{equation}
uniformly in $\gamma ,t\geq 0$.

Following Picard's method, let us replace $F$ in (\ref{ode}) by a locally
H\"{o}lder continuous function $f:[0,T]\longrightarrow \mathcal{B}$: 
\begin{equation*}
\left\| f(r)-f(s)\right\| \leq C\left| r-s\right| ^{\theta }
\end{equation*}
for $0\leq r\leq s<T$ and $\theta >0$. In this case, a solution to (\ref{ode})
is given by the variation of constants formula  
\begin{equation}
z(t)=e^{-tA}z_{0}+\int_{0}^{t}\,e^{-\left( t-s\right) A}\,f(s)\,ds\,.
\label{cvm}
\end{equation}
Note that $z:[0,T)\longrightarrow \mathcal{B}$ is continuously
differentiable with $z\in D(A)$ satisfying the differential equation (\ref
{ode}). Moreover, $z (t)$ is the unique solution with $z(0)=z_{0}$ provided $
f$ is such that $\displaystyle\lim _{\rho \to 0} \int _{0}^{\rho } \left\| f
(s)\right\| ds =0 $.

Now, substituting $f(s)=F\left( z(s)\right) $ into (\ref{cvm}) leads to an
integral equation 
\begin{equation}
z(t)=e^{-tA}z_{0}+\int_{0}^{t}\,e^{-(t-s)A}\,F\left( z(s)\right) \,ds
\label{integral}
\end{equation}
whose solution, whether it exists, also solves the initial value problem
(\ref{dpde}) provided $F\left( z(s)\right) $ is shown to be locally 
H\"{o}lder continuous on the interval $0\leq t<T$.

To formulate the necessary condition on $F$ and state our results, let $
\mathcal{B}^{\gamma }=D(A^{\gamma })$, $\gamma \geq 0$, denote the Banach
space with the graph norm 
\begin{equation*}
\left\| f\right\| _{\gamma }:=\left\| A^{\gamma }f\right\| .
\end{equation*}
$F:\mathcal{B}^{\gamma }\longrightarrow L_{\mathrm{p,o}}^{2}\left( -\pi ,\pi
\right) $ is said to be locally Lipschtzian if there exist $U\subset 
\mathcal{B} ^{\gamma }$ and a finite constant $L$ such that 
\begin{equation}
\left\| F(z_{1})-F(z_{2})\right\| \leq L\left\| z_{1}-z_{2}\right\| _{\gamma
}\,  \label{F-F}
\end{equation}
holds for any $z_{1},\,z_{2}\in U$.

\begin{theorem}
\label{thivp} The initial value problem (\ref{ode}) has a unique solution $
z(t)$ for all $t\in \mathbb{R}_{+}$ with $z(0)=z_{0}\in \mathcal{B}^{1/2}$.
In addition, if $\left\| z(t)\right\| _{1/2}$ is bounded as $t\rightarrow
\infty $, the trajectories $\left\{ z(t)\right\} _{t\geq 0}$ lie on a
compact set in $\mathcal{B}^{1/2}$.
\end{theorem}

\noindent \textbf{Proof.} The proof of Theorem \ref{thivp} will be divided
into four parts. Firstly, $F(z(t))$ will be shown to be H\"{o}lder
continuous under the Lipschtzian condition (\ref{F-F}), which establishes
the equivalence between the integral equation (\ref{integral}) and the
initial problem (\ref{ode}). Secondly, the Banach fixed point theorem will
be used to show the existence of a unique solution $z(t)$ of (\ref{integral})
for $0\leq t\leq T$. Hence, by a compactness argument, the solution $z(t)$ 
will be extended to all $t\in \mathbb{R}_{+}$. Finally, assuming that $
\left\| z(t)\right\| _{1/2}$ stays bounded for all $t>0$, we conclude the
proof. We have to wait till Section \ref{Stability} for the boundedness
hypothesis to be established.

\bigskip

\noindent \textit{Part I: Continuity.} Let us show that $F:D(A^{1/2})
\longrightarrow L_{\mathrm{o,p}}^{2}\left( -\pi ,\pi \right) $ given by $
F(z)=-2\alpha z\,z^{\prime }$ is locally Lipschitz. We note that $
D(A^{1/2})=H_{\mathrm{o,p}}^{1}\left( -\pi ,\pi \right) $ where $H_{\mathrm{
o,p}}^{k}\left( -\pi ,\pi \right) $ is the Sobolev space of odd periodic
functions which have distributional derivatives up to order $k$. It thus
follows that, if $z\in H_{\mathrm{o,p}}^{1}$, then $z(x)=\int_{0}^{x}z^{
\prime }(\xi )\,d\xi $ is absolutely continuous with 
\begin{equation*}
\sup_{x\in \lbrack -\pi ,\pi ]}\left| z(x)\right| \leq \sqrt{2\pi }\left\|
z\right\| _{1/2}\,,
\end{equation*}
by the Schwarz inequality. Moreover, using (\ref{b1}), we have 
\begin{eqnarray}
\left\| F(z_{1})-F(z_{2})\right\| &\leq &2\alpha \left\{ \left\|
z_{1}(z_{1}^{\prime }-z_{2}^{\prime })\right\| +\left\|
(z_{1}-z_{2})z_{2}^{\prime }\right\| \right\}  \label{lipsch} \\
&\leq &2\alpha \sqrt{2\pi }\left\{ \left\| z_{1}\right\| \left\|
z_{1}-z_{2}\right\| _{1/2}+\left\| z_{1}-z_{2}\right\| \left\| z_{2}\right\|
_{1/2}\right\}  \notag
\end{eqnarray}
which satisfies (\ref{F-F}) with $\gamma =1/2$ and $L=2\alpha \sqrt{2\pi }
\left( \left\| z_{1}\right\| _{1/2}+\left\| z_{2}\right\| _{1/2}\right) $.

Suppose that $z:(0,T)\longrightarrow \mathcal{B}^{1/2}$ is a continuous
solution of (\ref{integral}). From the estimate (\ref{b1}), we have 
\begin{eqnarray}
\left\| \left( e^{-hA}-I\right) e^{-\tau A}w\right\| _{1/2} &\leq
&\int_{0}^{h}\left\| A\,e^{-(s+\tau )A}w\right\| _{1/2}\,\,ds  \notag \\
&=&\int_{0}^{h}\left\| A^{1-\delta }\,e^{-sA}\right\| ds\,\left\| A^{\delta
}e^{-\tau A}w\right\| _{1/2}  \label{contbound} \\
&\leq &C_{1-\delta }\int_{0}^{h}\frac{1}{s^{1-\delta }}ds\,\left\| A^{\delta
}e^{-\tau A}w\right\| _{1/2}  \notag \\
&\leq &\frac{C_{1-\delta }}{\delta \,}h^{\delta }C_{\delta +1/2}\,\frac{
e^{-c\tau }}{\tau ^{\delta +1/2}}\,\left\| w\right\|  \notag
\end{eqnarray}
for $0<\delta <1/2$ which can be used in the equation (\ref{integral}) along
with (\ref{F-F}), to get 
\begin{equation}
\begin{array}{ccc}
\left\| z(t+h)-z(t)\right\| _{1/2} & \leq & \left\| \left( e^{-hA}-I\right)
e^{-tA}z_{0}\right\| _{1/2}+\int_{0}^{t}\left\| \left( e^{-hA}-I\right)
e^{-\left( t-s\right) A}F(z(s))\right\| _{1/2}\,ds \\ 
&  &  \\ 
&  & +\int_{t}^{t+h}\left\| e^{-\left( t+h-s\right) A}F(z(s))\right\|
_{1/2}\,ds\leq K\,h^{\delta }
\end{array}
\label{continuous}
\end{equation}
for some constant $K<\infty $ in the open interval $\left( 0,T\right) $.
Combined with (\ref{F-F}), this implies the H\"{o}lder continuity of $
f(t)=F\left( z(t)\right) $ and the equivalence between the equations
(\ref{ode}) and (\ref{integral}). 

\medskip

\noindent \textit{Part II: Local existence.} Let $V=\left\{ z\in \mathcal{B}
^{1/2}:\left\| z-z_{0}\right\| \leq \varepsilon \right\} $ be an $%
\varepsilon $--neighborhood and let $L$ be the Lipschitz constant of $F$ on $
V$. We set $B=\left\| F(z_{0})\right\| $ and let $T$ be a positive number
such that

\begin{equation}
\left\| \left( e^{-hA}-I\right) \,z_{0}\right\| _{1/2}\leq \frac{\varepsilon 
}{2}  \label{H1}
\end{equation}
with $0\leq h\leq T$ and 
\begin{equation}
C_{1/2}\left( B+L\varepsilon \right) \int_{0}^{T}s^{-1/2}\,e^{-cs}\,ds\leq 
\frac{\varepsilon }{2}  \label{H2}
\end{equation}
hold.

Let $\mathcal{S}$ denote the set of continuous functions $%
y:[t_{0},t_{0}+T]\longrightarrow \mathcal{B}^{1/2}$ such that $\left\|
y(t)-z_{0}\right\| \leq \varepsilon $. Equipped with the sup--norm 
\begin{equation*}
\left\| y\right\| _{T}:=\sup_{t_{0}\leq t\leq t_{0}+T}\left\| y(t)\right\|
_{1/2}
\end{equation*}
$\mathcal{S}$ is a complete metric space.

Defining $\Phi \lbrack y]:[t_{0},t_{0}+T]\longrightarrow \mathcal{B}^{1/2}$
for each $y\in \mathcal{S}$ by 
\begin{equation*}
\Phi \lbrack y](t)=e^{-(t-t_{0})A}z_{0}+\int_{t_{0}}^{t}\,e^{-\left(
t-s\right) A}\,F\left( y(s\right) )\,\,ds\,,
\end{equation*}
we now show that, under the conditions (\ref{H1}) and (\ref{H2}), $\Phi : 
\mathcal{S}\longrightarrow \mathcal{S}$ is a strict contraction. Using 
\begin{equation*}
\left\| F(y(t))\right\| \leq \left\| F(y(t))-F(z_{0})\right\| +\left\|
F(z_{0})\right\| \leq L\left\| y(t)-z_{0}\right\| _{1/2}+B\leq L\varepsilon
+B
\end{equation*}
and (\ref{b1}), we have 
\begin{eqnarray*}
\left\| \Phi \lbrack y](t)-z_{0}\right\| _{1/2} &\leq &\left\| \left(
e^{-(t-t_{0})A}-I\right) e^{-tA}z_{0}\right\|
_{1/2}+\int_{t_{0}}^{t_{0}+T}\left\| A^{1/2}e^{-\left( t-s\right) A}\right\|
\,\left\| F(y(s))\right\| \,ds \\
&\leq &\frac{\varepsilon }{2}+C_{1/2}\left( B+L\varepsilon \right)
\int_{0}^{T}s^{-1/2}\,e^{-cs}\,ds\leq \varepsilon
\end{eqnarray*}
and since $\Phi \lbrack y]$ is continuous by an estimate analogous to
(\ref{continuous}), $\Phi \lbrack y]\in \mathcal{S}$. 

Analogously, from (\ref{F-F}) and (\ref{H2}), for any $y,w\in \mathcal{S}$ 
\begin{eqnarray*}
\left\| \Phi \lbrack y](t)-\Phi \lbrack w](t)\right\| _{1/2} &\leq
&\int_{t_{0}}^{t_{0}+T}\left\| A^{1/2}e^{-\left( t-s\right) A}\right\|
\,\left\| F(y(s))-F(w(s))\right\| \,ds \\
&\leq &C_{1/2}L\int_{0}^{T}s^{-1/2}\,e^{-cs}\,ds\,\left\| y-w\right\|
_{T}\leq \frac{1}{2}\left\| y-w\right\| _{T}
\end{eqnarray*}
holds uniformly in $t\in \lbrack t_{0},t_{0}+T]$ concluding our claim.

By the contraction mapping theorem, $\Phi $ has a \textbf{unique} fixed
point $z $ in $\mathcal{S}$ which is the continuous solution of the integral
equation (\ref{integral}) on $(t_{0},t_{0}+T)$ and, by \textit{Part I}, is
the solution of (\ref{ode}) in the same interval with $z(t_{0})=z_{0}\in 
\mathcal{B}^{1/2}$.

\medskip

\noindent \textit{Part III: Global existence.} As the set $U$ where (\ref
{F-F}) holds is compact, the same $T$ can be chosen in \textit{Part II} for
any initial condition $z_{0}\in U$. Moreover, if $I_{1}=(t_{1},t_{1}+T)$ and 
$I_{2}=(t_{2},t_{2}+T)$ are two intervals containing $t_{0}$, then there
exist $z_{0,1},z_{0,2}\,\in U$ such that the two solutions $z_{1}(t)$ and $
z_{2}(t)$ of equation (\ref{ode}) on $I_{1}$ with $z_{1}(t_{1})=z_{0,1}$ and
on $I_{2}$ with $z_{2}(t_{2})=z_{0,2}$, respectively, coincide in the open
interval $I_{1}\cap I_{2}$. As a consequence, one can define an \textbf{open
maximal interval} $I_{\mathrm{\max }}=(t_{-},t_{+})$ (containing the
origin), where the solution $z(t)$ of (\ref{ode}) is uniquely given by
patching together the solutions $z_{j}(t)$ on intervals $I_{j}$ with $
z_{j}(t_{j})=z_{0,j}$. By construction, there is no solution to (\ref{ode})
on $(t_{0},t^{\prime })$ if $t^{\prime }>t_{+}$. Therefore, either $
t_{+}=\infty $, or else there exist a sequence $\left\{ t_{n}\right\} _{n\in 
\mathbb{N}_{+}}$, with $t_{n}\rightarrow t_{+}$ as $n\rightarrow \infty $
such that $z(t_{n})$ tend to the boundary $\partial U$ of the compact set $U$
.

It thus follows that, if $t_{+}$ is finite, the solution $z(t)$ blows--up at
finite time. In what follows we show that $\left\| z(t)\right\| _{1/2}$
remains finite for all $t>t_{0}$ and this implies global existence of $z(t)$
. Let us start with the following generalization of the Gronwall inequality.

\begin{lemma}[Gronwall]
\label{gronwall}Let $\xi $ and $\gamma $ be numbers and let $\theta $ and $
\zeta $ be non--negative continuous functions defined in a interval $
I=\left( 0,T\right) $ such that $\xi \geq 0$, $\gamma >0$ and 
\begin{equation}
\zeta (t)\leq \theta (t)+\xi \int_{0}^{t}\left( t-\tau \right) ^{\gamma
-1}\,\zeta (\tau )\,d\tau \,.  \label{zeta}
\end{equation}
Then 
\begin{equation}
\zeta (t)\leq \theta (t)+\int_{0}^{t}E_{\gamma }^{\prime }(t-\tau )\,\theta
(\tau )\,d\tau   \label{zeta1}
\end{equation}
holds for $t\in I$, where $E_{\gamma }^{\prime }=dE_{\gamma }/dt$, 
\begin{equation}
E_{\gamma }(t)=\sum_{n=0}^{\infty }\frac{1}{\Gamma \left( n\gamma +1\right) }
\left( \xi \Gamma (\gamma )\,t^{\gamma }\right) ^{n}  \label{Egamma}
\end{equation}
and $\Gamma (z)=\displaystyle\int_{0}^{\infty }t^{z-1}e^{-t}dt$ is the gamma
function. In addition, if $\theta (t)\leq K$ for all $t\in I$, then 
\begin{equation}
\zeta (t)\leq K\,E_{\gamma }(t)\,\leq K^{\prime }\,e^{\xi \Gamma (\gamma )T}
\label{zeta2}
\end{equation}
holds for some finite constant $K^{\prime }$.
\end{lemma}

\noindent \textbf{Proof.} If $\mathcal{T}$ \ is an integral operator given
by the convolution 
\begin{equation}
\mathcal{T}\zeta (t)=\xi \int_{0}^{t}\left( t-\tau \right) ^{\gamma -1}\zeta
(\tau )\,d\tau \,,  \label{tau}
\end{equation}
then the inequality (\ref{zeta}) can be formally solved by 
\begin{equation*}
\zeta (t)=\theta (t)+\sum_{n=1}^{\infty }\mathcal{T}^{n}\theta (t)
\end{equation*}
where $\mathcal{T}^{n}$ is also an convolution integral operator which can
be explicitly evaluated by the Laplace transform, 
\begin{eqnarray*}
\mathcal{T}^{n}\theta (t) &=&\frac{1}{\Gamma \left( n\gamma \right) }\left(
\xi \Gamma (\gamma )\right) ^{n}\int_{0}^{t}\left( t-\tau \right) ^{n\gamma
-1}\theta (\tau )\,d\tau \\
&=&\frac{1}{\Gamma \left( n\gamma +1\right) }\left( \xi \Gamma (\gamma
)\right) ^{n}\int_{0}^{t}\frac{d}{dt}\left( t-\tau \right) ^{n\gamma }\theta
(\tau )\,d\tau \equiv \left( f_{n}^{\prime }\ast \theta \right) (t)\,,
\end{eqnarray*}
with $f_{n}(t)=\left( \xi \Gamma (\gamma )\,t^{\gamma }\right) ^{n}/\Gamma
\left( n\gamma +1\right) $.

Equation (\ref{zeta1}) (and (\ref{zeta2}) by the fundamental theorem of
calculus) thus follows by setting $E_{\gamma }(t)=\sum_{n\in \mathbb{N}
}f_{n}(t)$. Note that this series is absolutely and uniformly convergent in $
t\in I$, with $E_{\gamma }(0)=1$, and it cannot grow faster than exponential 
\begin{equation}
E_{\gamma }(T)\sim \frac{1}{\gamma }e^{\xi \Gamma (\gamma )T}  \label{exp}
\end{equation}
as $T\rightarrow \infty $ (see Lemma $7.1.1$ in \cite{H}). This concludes
the proof of Lemma \ref{gronwall}.

\hfill $\Box $

Taking the graph norm of (\ref{integral}), we have in view of (\ref{bounds}),
(\ref{b1}) and (\ref{exp})  
\begin{eqnarray}
\left\| z(t)\right\| _{1/2} &\leq &\left\| e^{-(t-t_{0})A}z_{0}\right\|
_{1/2}+L\int_{t_{0}}^{t}\,\left\| A^{1/2}e^{-(t-s)A}\right\| \,\left\|
\,z(s)\right\| _{1/2}\,\,ds  \notag \\
&\leq &C\left\| z_{0}\right\| _{1/2}+L\int_{t_{0}}^{t}\,\left( t-s\right)
^{-1/2}\,\left\| \,z(s)\right\| _{1/2}\,\,ds  \label{b2} \\
&\leq &C\,\exp \left( LC_{1/2}\sqrt{\pi }t\right) \,\left\| z_{0}\right\|
_{1/2}\,,  \notag
\end{eqnarray}
which is finite for any $t\in \mathbb{R}_{+}$.

\textit{Part IV: Compact trajectories.} Since $\mathcal{B}^{\gamma }\subset 
\mathcal{B }^{1/2}$ has compact inclusion if $1/2<\gamma <1$ \cite{H}, it
suffices to show that $\left\| z(t)\right\| _{\gamma }$ remains bounded as $
t\rightarrow \infty $. The hypothesis $\left\| z(t)\right\| _{1/2}<\infty $
combined with (\ref{lipsch}) implies the existence of $C^{\prime }<\infty $
such that, analogously as in (\ref{b2}), 
\begin{eqnarray*}
\left\| z(t)\right\| _{\gamma } &\leq &\left\| e^{-tA}z_{0}\right\| _{\gamma
}+\int_{0}^{t}\,\left\| A^{\gamma }e^{-(t-s)A}\right\| \,\left\| F\left(
\,z(s)\right) \right\| \,\,ds \\
&\leq &C_{\gamma -1/2}\,t^{1/2-\gamma }\,e^{-ct}\left\| z_{0}\right\|
_{1/2}+C^{\prime }\,C_{\gamma }\int_{0}^{t}\,\left( t-s\right) ^{-\gamma
}\,e^{-c(t-s)}\,\,ds\,,
\end{eqnarray*}
which is bounded for $t>0$ provided $c>0$ (i.e. $\inf_{\lambda }\sigma (A)>0$
). Although the spectrum of $A$ is not positive if $\beta \leq 8\pi ,$ we
shall see in Section \ref{Stability} that $A$ in the integral equation
(\ref{integral}) can be replaced by a positive linear operator $L$ (see
Theorems \ref{asym} and \ref{positive}).

This concludes the proof of Theorem \ref{thivp}.

\hfill $\Box $

It follows by analogous procedure that if $z_{1}$ and $z_{2}$ are solutions
of (\ref{ode}) differing by their initial value in $\mathcal{B}^{1/2}$,\ then

\begin{eqnarray*}
\left\| z_{1}(t)-z_{2}(t)\right\| _{1/2} &\leq &\left\| e^{-tA}\left(
z_{0,1}-z_{0,2}\right) \right\| _{1/2}+\int_{0}^{t}\left\| A^{1/2}e^{-\left(
t-s\right) A}\right\| \,\left\| F(z_{1}(s))-F(z_{2}(s))\right\| \,ds \\
&\leq &\left\| e^{-tA}\left( z_{0,1}-z_{0,2}\right) \right\|
_{1/2}+C_{1/2}L\int_{0}^{t}\left( t-s\right) ^{-1/2}\,e^{-cs}\,ds\,\left\|
z_{1}(s)-z_{2}(s)\right\| _{1/2}
\end{eqnarray*}
which implies, by the Gronwall inequality, the continuous dependence of $
z(t) $ with respect to its initial condition.

We may also consider the dependence of $z$ with respect to the parameter $
\alpha =\beta /(4\pi )$. The next statement is a corollary of the above
analysis.

\begin{theorem}
\label{cdi} The solution $z(t):\mathbb{R}_{+}\times \mathcal{B}%
^{1/2}\longrightarrow \mathcal{B}^{1/2}$ to the initial value problem
(\ref{ode}) as a function of the bifurcation parameter $\alpha $ and the
initial value $z_{0}$ is continuous.
\end{theorem}

\begin{remark}
\label{DA} It can be shown (see \cite{H}) that for any initial value $
z_{0}\in \mathcal{B}^{\gamma }$, $0<\gamma <1$, the solution is actually in $
D(A)$ at any later time. Moreover, since $F:\mathcal{B}^{1/2}\longrightarrow
L_{\mathrm{o,p}}^{2}\left( -\pi ,\pi \right) $ is $\mathcal{C}^{\infty }$
(has Fr\'{e}chet derivatives of all orders), it can also be shown that $
\left( \alpha ,z_{0}\right) \in \mathbb{R}_{+}\times \mathcal{B}
^{1/2}\longrightarrow z(t;\alpha ,z_{0})$ is $\mathcal{C}^{\infty }$ \ for
all $t>0$.
\end{remark}

\begin{remark}
\label{sobolev} Under minor modifications, one can show existence,
uniqueness and continuous dependence of (\ref{dpde}) in Sobolev space $H_{
\mathrm{o,p}}^{1}\left( -\pi ,\pi \right) $ with norm $\left\| z\right\|
_{1}=\left\| z^{\prime }\right\| _{L_{\mathrm{o,p}}^{2}}$ (just include the
linear term of (\ref{dpde}) in the definition of $F$). The same results hold
for equation (\ref{pde}) in the Sobolev space of even and periodic function $
H_{\mathrm{e,p}}^{1}\left( -\pi ,\pi \right) $ with both norms $\left\|
\cdot \right\| _{1}$ and $\left\| \cdot \right\| _{1/2}$. Note from item $2.$
after (\ref{bounds}) and (\ref{A}) that $\alpha \left\| z\right\|
_{1}=\left\| z\right\| _{1/2}+2\left\| z\right\| _{L_{\mathrm{o,p}}^{2}}$ so,
both norms are equivalent.
\end{remark}


\section{Equilibrium Solutions \label{SS}}

\setcounter{equation}{0} \setcounter{theorem}{0}

Time independent (equilibrium) solutions of (\ref{dpde}) are odd solutions
of the ordinary differential equation 
\begin{equation}
\alpha \left( \psi ^{\prime \prime }-2\psi \psi ^{\prime }\right) +2\psi
=0\,,  \label{stationary}
\end{equation}
with periodic conditions $\psi (-\pi )=\psi (\pi )$ and $\psi ^{\prime
}(-\pi )=\psi ^{\prime }(\pi )$, $\alpha =\beta /\left( 4\pi \right) \geq 0$
, which can be written as 
\begin{equation}
\left\{ 
\begin{array}{lll}
w^{\prime } & = & 2p\left( w-\alpha ^{-1}\right) \\ 
&  &  \\ 
p^{\prime } & = & w\,,
\end{array}
\right.  \label{syst}
\end{equation}
by setting $p=\psi $ and $w=\psi ^{\prime }$.

In this section we give a qualitative and quantitative description of the
solutions of (\ref{syst}) in the phase space $\mathbb{R}^{2}$ and study
their implications for the equilibrium solutions of (\ref{dpde}). Our
results are summarized as follows.

\begin{theorem}
\label{existence} The stationary equation (\ref{stationary}) has two
distinct regimes separated by $\alpha =2$ ($\beta =8\pi $). For $\alpha \geq
2$, $\psi _{0}\equiv 0$ is the unique solution. For $\alpha <2$ such that $
2/\left( k+1\right) ^{2}\leq \alpha <2/k^{2}$ holds for some $k\in \mathbb{N}
_{+}$, there exist $2k$ non--trivial solutions $\psi _{j}^{+},\psi _{j}^{-}$
, $j=1,\ldots ,k$, with fundamental period $2\pi /j$, $\psi _{j}^{\pm
}(-x)=-\psi _{j}^{\pm }(x)$ and $\psi _{j}^{-}(x)=\psi _{j}^{+}(x+\pi )$.
Moreover, each pair of non--trivial solutions bifurcate from the trivial
solution $\psi _{0}$ at $\alpha _{j}=2/j^{2}$ ($\beta _{j}=8\pi /j^{2}$)
with $\lim\limits_{\alpha \uparrow \alpha _{j}}\psi _{j}^{\pm }=0$.

In the phase space, these solutions $\left( \psi _{j}^{\prime },\psi
_{j}\right) $, are closed orbits around $\left( 0,0\right) $ whose distance
from the origin increases monotonically as $\alpha $ decreases. Numerical
computations indicate that these orbits approach rapidly to the open orbit $
\left\{ \left( \alpha ^{-1},\alpha ^{-1}x\right) ,\,x\in \mathbb{R}\right\} $
from the left as $\alpha \rightarrow 0$.
\end{theorem}

Let us begin by stating the general properties derived by the same tools
used in the analysis performed in Section \ref{EU}.

The vector field $f:\mathbb{R}^{2}\longrightarrow \mathbb{R}^{2}$, 
\begin{equation*}
(w,p)\longrightarrow f(w,p)=\left( 2p(w-\alpha ^{-1}),w\right) \,,
\end{equation*}
in the right hand side of (\ref{syst}), defines a smooth autonomous
dynamical system. It thus follows from Piccard's theorem (see e.g. \cite{CL})
that there exist a unique solution $(w(x),p(x))$ of this system, globally 
defined in $\mathbb{R}^{2}$, with $(w(0),p(0))=(w_{0},p_{0})$. As we have
seen in Section \ref{EU}, the existence of a global solution and its
continuous dependence on the value $(w_{0},p_{0})$, and on the parameter $
\alpha $, follow from Gronwall's lemma, which holds here in its standard
form. As a consequence, the phase space $\mathbb{R}^{2}$ is foliated by
non--overlapping orbits 
\begin{equation*}
\gamma _{P}=\left\{ (w(x),h(x)):x\in \mathbb{R\;}\mathrm{and}\;P=(w(0),p(0))
\right\}
\end{equation*}
which passes by $P=(w_{0},p_{0})\in \mathbb{R}^{2}$ at $x=0$. Note that, by
varying continuously $P$ and $\alpha $, the orbit $\gamma _{P}$ varies
continuously in the phase space.

We shall now determine the values $(P,\alpha )$ by which the solution of
(\ref{syst}) defines closed orbits. Note that the orbits are symmetric with 
respect to the $w$--axis, $L=\{(w,0):w\in \mathbb{R}\}$, since the system of
equations (\ref{syst}) remains invariant if the sign of both, $x$ and $p$,
are reversed. As we shall see, there is no loss of generality if the initial
value $(w(0),p(0))=P$ belongs to $L$. We write $\gamma _{P}=\gamma _{w_{0}}$.

\begin{proposition}
\label{orbit} Every orbit $\gamma _{P}$ is determined by a single value $P$
in the positive semi--axis $L^{+}=\{(w_{0},0):w_{0}\geq 0\}$. For $w_{0}>0$,
the orbit $\gamma _{w_{0}}$ is either closed or unbounded depending on
whether $\alpha \,w_{0}<1$ or $\alpha \,w_{0}\geq 1$, respectively. The
orbit $\gamma _{\alpha ^{-1}}=\{(\alpha ^{-1},\alpha ^{-1}x):x\in \mathbb{R}
\}$ separates the phase space $\mathbb{R}^{2}$ in such way that $\gamma _{P}$
is closed if $P$ is on the left of $\gamma _{\alpha ^{-1}}$ and unbounded
otherwise. In addition, if $w_{0}=0$, then $\gamma _{0}=\{(0,0)\}$, and the
origin is enclosed by every closed orbit.
\end{proposition}

\noindent \textbf{Proof.} The proof of Proposition \ref{orbit} follows from
an explicit computation. By the chain rule, equation (\ref{syst}) can be
written as 
\begin{equation}
\frac{dp}{dw}=\frac{w}{2p\left( w-\alpha ^{-1}\right) }  \label{dh/dw}
\end{equation}
provided $\alpha w\neq 1$. The trajectories $\gamma _{w_{0}}$, obtained by
integrating $2p\,dp=w\,dw/\left( w-\alpha ^{-1}\right) $ with initial point $
P=(w_{0},0)$, 
\begin{equation}
p^{2}=w-w_{0}+\alpha ^{-1}\ln \left( \frac{1-\alpha w}{1-\alpha w_{0}}
\right) \,,  \label{h2}
\end{equation}
are portrayed in Figure $1$.


\begin{figure}[!ht] 
\begin{center}
\epsfig{file=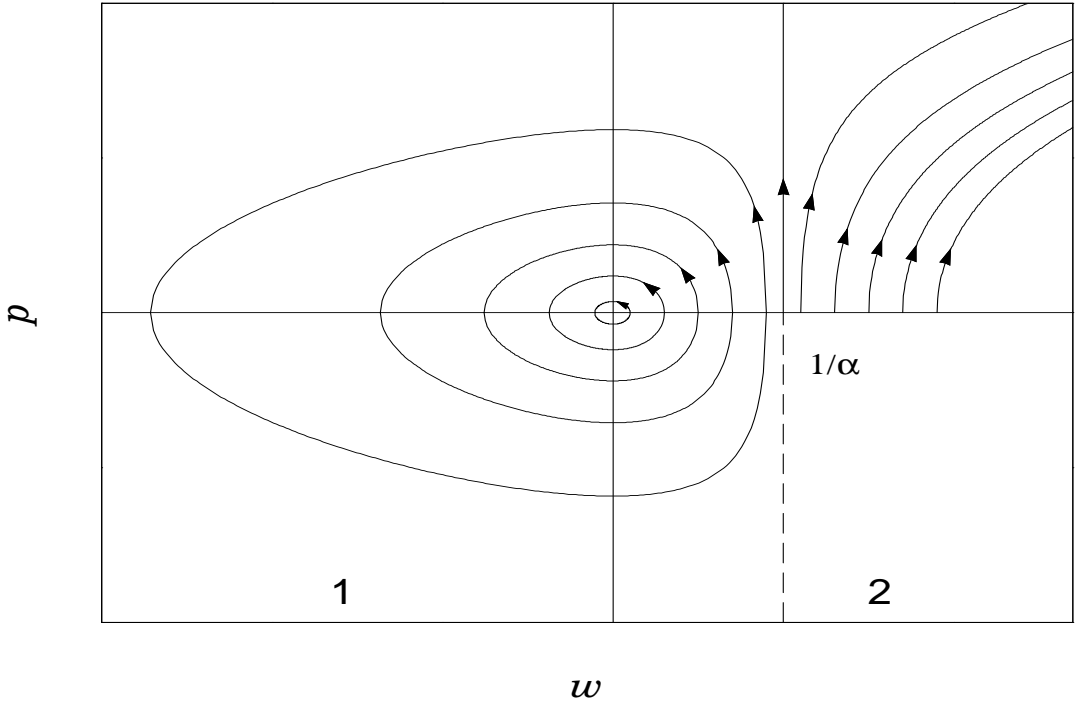, height=9.5cm, width=14.5cm}
\end{center}
\caption{Trajectories of the dynamical system (\ref{syst}).} 
\label{phsp} 
\end{figure}


We note that $P=(0,0)$ is the only critical point of (\ref{syst}) which is a
center for all $\alpha >0$ since, by linearizing $f(w,p)$ around $P=(0,0)$
gives a matrix whose eigenvalues are $\lambda _{\pm }=\pm i\sqrt{2\alpha
^{-1}}$. This implies that $\gamma _{0}=\{(0,0)\}$ and the orbits $\gamma
_{w_{0}}$ with $w_{0}$ sufficiently closed to $0$ are, in view of (\ref{h2}),
ellipses defined by the equation $2\alpha ^{-1}p^{2}+w^{2}=C$. 

When $\alpha w_{0}=1$, using mathematical induction and equations (\ref{syst})
with $(w(0),p(0))=(w_{0},0)$, we have  
\begin{equation*}
\frac{d^{n}w}{dx^{n}}(0)=0\,,
\end{equation*}
for all $n\geq 1$, which leads 
\begin{equation*}
\gamma _{\alpha ^{-1}}=\left\{ \left( \alpha ^{-1},\alpha ^{-1}x\right)
:x\in \mathbb{R}\right\} \,.
\end{equation*}

Hence, if $\omega =\omega (P)$ denotes the set of limit points (the $\omega $
--limit set) given by 
\begin{equation}
\omega (P)=\left\{ (w^{\ast },h^{\ast })\in \mathbb{R}^{2}:\lim_{n%
\rightarrow \infty }\left( w(x_{n}),h(x_{n})\right) =(w^{\ast },h^{\ast
})\right\}  \label{omega}
\end{equation}
for some sequence of points $\{x_{n}\}$ such that $x_{n}\rightarrow \infty $
, as $n\rightarrow \infty $, $\gamma _{\alpha ^{-1}}$ separates two
different type of orbits: $\omega (P)=\gamma _{P}$ or $\omega (P)=\{\infty
\} $ depending on whether the point $P$ is at the left or at the right of $
\gamma _{\alpha ^{-1}}$.\hfill

\hfill $\Box $

\smallskip

\noindent \textbf{Proof of Theorem \ref{existence}.} The stationary
solutions satisfy (\ref{syst}) with periodic \ conditions $w(0)=w(2\pi )$
and $p(0)=p(2\pi )$. By fixing the period $T$ of an orbit $\gamma _{w_{0}}$
in $2\pi $, the label $w_{0}$ becomes implicitly dependent on the parameter $
\alpha $. In view of Proposition \ref{orbit}, Theorem \ref{existence}
follows if for $\alpha \geq 2$, except by the orbit $\gamma _{0}=\{(0,0)\}$,
no (non--trivial) solution has period $T=2\pi $ and for $\alpha <2$ there is
a one--to--one correspondence between $w_{0}$ and $\alpha $ for $T$ fixed
at\ any value $2\pi /k$, $k=1,\ldots ,\left[ \sqrt{2/\alpha }\right] $.

More precisely, let $T=T(\alpha ,w_{0})$ denote the period of the dynamical
system (\ref{syst}) with initial value $(w(0),p(0))=\left( w_{0},0\right) $: 
\begin{equation}
T=\int_{\gamma _{w_{0}}}dx=2\int \frac{dp}{w}\,,  \label{T}
\end{equation}
where, by symmetry, the second integration is over the semi--orbit above the 
$w$--axis. For $\mathcal{D}=\left\{ \left( \alpha ,w_{0}\right) \in
  \mathbb{R}_{+}\times \mathbb{R}_{+}:\alpha w_{0}\leq 1\right\} $, we set  
\begin{equation*}
G_{j}=T-\frac{2\pi }{j}
\end{equation*}
and note that $G_{j}:\mathcal{D}\longrightarrow \mathbb{R}$ is a continuous
function of both variables satisfying 
\begin{equation}
G_{j}\left( 2/j^{2},0\right) =0\text{.}  \label{Gj}
\end{equation}

To see (\ref{Gj}), we compute the period $T_{L}$ of an elliptic orbit, e.g. $
\left\{ \left( 2/\alpha \right) p^{2}+w^{2}=1\right\} $, of (\ref{syst})
linearized at the origin ($f(w,p)$ replaced by $\left( 2\alpha
^{-1}p,w\right) $), 
\begin{equation}
T_{L}=4\int_{0}^{\sqrt{\alpha /2}}\frac{dp}{\sqrt{1-\left( 2/\alpha \right)
p^{2}}}=2\pi \sqrt{\frac{\alpha }{2}\,}\,,  \label{TL}
\end{equation}
and note that $\lim_{w_{0}\rightarrow 0}T(\alpha ,w_{0})=T_{L}$. Continuity
follows from the general properties stated previously.

Hence, provided 
\begin{equation}
\frac{\partial T}{\partial w_{0}}>0  \label{dtdw0}
\end{equation}
holds for all $\left( \alpha ,w_{0}\right) \in \mathcal{D}$, by the implicit
function theorem, there exists a \textbf{unique} (strictly) monotone
decreasing function $\widehat{w}_{j}:\left[ 0,2/j^{2}\right] \longrightarrow 
\mathbb{R}_{+}$ with $\widehat{w}_{j}(2/j^{2})=0$ such that $G_{j}(\alpha ,
\widehat{w}_{j}(\alpha ))=0$. Note that (\ref{dtdw0}) and 
\begin{equation}
T(\alpha ,w_{0})=\sqrt{\alpha }T(1,\alpha w_{0})\,  \label{TT}
\end{equation}
imply that $T$ is an increasing function of both $\alpha $ and $w_{0}$,
independently. This fact, which can be seen by rescaling (\ref{syst}) by $
x\rightarrow \overline{x}=x/\sqrt{\alpha }$, $w\rightarrow \overline{w}
=\alpha w$ and $p\rightarrow \overline{p}=\sqrt{\alpha }p$, explains the
monotone behavior of $\widehat{w}_{j}$.

It thus follows that, if $\alpha <2$, for each $j=1,\ldots ,k$ such that $
2/\left( k+1\right) ^{2}\leq \alpha <2/k^{2}$ holds, a unique function $
\widehat{w}_{j}$ such that $\widehat{w}_{j}(2/j^{2})=0$ exists. The
non--trivial solutions $\psi _{1}^{\pm },\ldots ,\psi _{k}^{\pm }$ of
(\ref{stationary}) are the $p$--component of $\gamma _{\widehat{w}_{j}}$, $
j=1,\ldots ,k$, which winds around the origin $j$--times: $\psi _{j}^{+}$ is 
$2\pi $--periodic odd function with fundamental period $2\pi /j$, $\left(
\psi _{j}^{+}\right) ^{\prime }(0)>0$ and satisfies $\psi _{j}^{+}(x+\pi
)=\psi _{j}^{-}(x)$. If $\alpha \geq 2$, because $T(\alpha ,w_{0})$ is a
strictly increasing function of $w_{0}$ and $T(\alpha ,0)\geq 2\pi $ (see
eq. (\ref{TL})), there is no solution of $G_{j}(\alpha ,w_{0})=0$ besides $
\widehat{w}_{j}(\alpha )=0$ for $j=1$. This reduces the proof of Theorem
\ref{existence} to the proof of inequality (\ref{dtdw0}). 

To prove (\ref{dtdw0}), it is convenient to change variables. Let 
\begin{equation}
q=\ln \left( 1-\alpha \,w\right) \,  \label{phi}
\end{equation}
be defined for $\alpha w<1$. From (\ref{TT}), there is no loss of generality
in taking $\alpha =1$. The system of equations (\ref{syst}) under this
condition is thus equivalent to the following Hamiltonian system\footnote{
We thank G. Benfatto for explaining this tranformation and for pointing us
equation (\ref{h2}) in a footnote of \cite{F}.} 
\begin{equation}
\left\{ 
\begin{array}{lll}
q^{\prime } & = & 2p \\ 
&  &  \\ 
p^{\prime } & = & 1-e^{q},
\end{array}
\right.  \label{syst1}
\end{equation}
whose energy function is given by 
\begin{equation}
H(q,p)=p^{2}+e^{q}-q-1\,.  \label{hamiltonian}
\end{equation}

The trajectory equation (\ref{h2}), when written in terms of the $q$%
--variable, \ gives exactly the energy level equation $H(q,p)=E$ with 
\begin{equation}
E=-w_{0}-\ln \left( 1-w_{0}\right) \,.  \label{Ew0}
\end{equation}
We denote by $\gamma _{E}$ the orbits of (\ref{syst1}) and note that, in
view of the fact 
\begin{equation*}
\frac{dE}{dw_{0}}=\frac{w_{0}}{1-w_{0}}\,>0,
\end{equation*}
there is a one--to--one correspondence between the two families of closed
orbits $\left\{ \gamma _{w_{0}},\,0\leq w_{0}<1\right\} $ and $\left\{
\gamma _{E},\,0\leq E<\infty \right\} $.

Now, let $\widetilde{T}=\widetilde{T}(E)$ be the period of an orbit $\gamma
_{E}$, 
\begin{equation}
\widetilde{T}=\int_{\gamma _{E}}dx=\int_{q_{-}}^{q_{+}}\frac{dq}{p}\,.
\label{Ttilde}
\end{equation}
Using the energy conservation law, we have 
\begin{equation}
p=p(q,E)=\sqrt{E-v(q)}\,,  \label{p}
\end{equation}
where the potential energy is given by 
\begin{equation}
v(q)=e^{q}-q-1\,,\,  \label{v}
\end{equation}
and $q_{\pm }=q_{\pm }(E)$ are the positive and negative roots of equation $%
v(q)=E$.

Equation (\ref{dtdw0}) holds if and only if $\dfrac{d\widetilde{T}}{dE}>0$
holds uniformly in $E\in \mathbb{R}_{+}$. But this follows from the
monotonicity criterion given by C. Chicone \cite{C} (see also \cite{CG}):

\begin{lemma}
\label{monotone}Let $v\in \mathcal{C}^{3}(\mathbb{R})$ be a three--times
differentiable function and let $f(q)=-v^{\prime }(q)$ be the force acting
at $q$. If $v/f^{2}$ is a convex function with 
\begin{equation}
\left( \frac{v}{f^{2}}\right) ^{\prime \prime }=\frac{6v\left( v^{\prime
\prime }\right) ^{2}-3\left( v^{\prime }\right) ^{2}v^{\prime \prime
}-2vv^{\prime }v^{\prime \prime \prime }}{\left( v^{\prime }\right) ^{4}}
>0\,,\qquad q\neq 0\,,  \label{convex}
\end{equation}
then the period $\widetilde{T}$ is a monotone (strictly) increasing function
of $E$.
\end{lemma}

\noindent \textbf{Proof.} It follows from (\ref{p}) two basic facts:

\begin{equation}
\frac{\partial p}{\partial q}=\frac{f}{2p}\hspace{1in}\mathrm{and\hspace{1in}%
} p(q_{\pm },E)=0\,.  \label{dphi}
\end{equation}
These will be used for deriving an appropriated integral representation of $%
d \widetilde{T}/dE$.

Let 
\begin{equation}
K:=\frac{1}{3}\int_{q_{-}}^{q_{+}}p^{3}\left( \frac{v}{f^{2}}\right)
^{\prime \prime }\,dq\,.  \label{K}
\end{equation}
Integrating twice by parts, gives 
\begin{eqnarray*}
K &=&\left. \frac{p^{3}}{3}\,\left( \frac{v}{f^{2}}\right) ^{\prime }\right|
_{q_{-}}^{q_{+}}-\left. \frac{pv}{2f}\right|
_{q_{-}}^{q_{+}}+\int_{q_{-}}^{q_{+}}\left( pf\right) ^{\prime }\,\frac{v}{
f^{2}}\,dq \\
&=&\frac{1}{2}\int_{q_{-}}^{q_{+}}\left( \frac{v}{2p}+vp\frac{f^{\prime }}{
f^{2}}\right) \,dq
\end{eqnarray*}
in view of (\ref{dphi}). Note that $f(q_{\pm })\not=0$ since 
\begin{equation*}
v^{\prime }(q_{\pm })=v(q_{\pm })-q_{\pm }=E-q_{\pm }
\end{equation*}
vanishes only at $E=0$. This follows from the fact that $v$ is a convex
positive function with $v(0)=0$ and asymptotic behavior $v(q)\sim q-1$ and $
\sim e^{\alpha q}$, as $q$ goes to $-\infty $ and $\infty $.

Now, using $(v/f)^{\prime }=v^{\prime }/f-vf^{\prime }/f^{2}=-1-vf^{\prime
}/f^{2}$, and integrating by parts, we continue 
\begin{eqnarray}
K &=&\frac{1}{2}\int_{q_{-}}^{q_{+}}\left( \frac{v}{2p}-p\left( \frac{v}{f}
\right) ^{\prime }-p\right) \,dq  \notag \\
&=&\frac{1}{2}\int_{q_{-}}^{q_{+}}\left( \frac{v}{p}-p\right) \,dq-\left. 
\frac{1}{2}p\,\left( \frac{v}{f}\right) \right| _{q_{-}}^{q_{+}}  \label{KK}
\\
&=&\frac{1}{2}\int_{q_{-}}^{q_{+}}\left( \frac{E}{p}-2p\right) \,dq  \notag
\end{eqnarray}
where in the last equation we have used $v=E-p^{2}$.

From (\ref{Ttilde}), (\ref{K}) and (\ref{KK}), we have 
\begin{equation*}
E\widetilde{T}=2\int_{q_{-}}^{q_{+}}p\,dq+\frac{2}{3}%
\int_{q_{-}}^{q_{+}}p^{3}\,\left( \frac{v}{f^{2}}\right) ^{\prime \prime
}\,dq\,.
\end{equation*}
Differentiating this with respect to $E$ and using (\ref{dphi}), gives 
\begin{equation*}
\widetilde{T}+E\,\frac{d\widetilde{T}}{dE}=\int_{q_{-}}^{q_{+}}\frac{dq}{p}
+\int_{q_{-}}^{q_{+}}p\,\left( \frac{v}{f^{2}}\right) ^{\prime \prime }\,dq
\end{equation*}
which, in view of (\ref{Ttilde}) and the assumption of Lemma \ref{convex},
implies 
\begin{equation*}
\frac{d\widetilde{T}}{dE}=\frac{1}{E}\int_{q_{-}}^{q_{+}}p\,\left( \frac{v}{
f^{2}}\right) ^{\prime \prime }\,dq>0\,.
\end{equation*}
\hfill

\hfill $\Box $

It remains to verify (\ref{convex}) for $v$ given by (\ref{v}). By an
explicit computation (see Chicone \cite{C}) 
\begin{equation*}
\left( \frac{v}{f^{2}}\right) ^{\prime \prime }\left( v^{\prime }\right)
^{4}=e^{q}\,g(q)
\end{equation*}
where 
\begin{equation*}
g(q):=e^{2q}+4\left( 1-q\right) e^{q}-2q-5
\end{equation*}
is such that $g(0)=g^{\prime }(0)=0$ and $g^{\prime \prime
}(q)=4e^{q}v(q)\geq 0$. This implies $g(q)\geq 0$ ($g(q)=0$ only if $q=0$),
the hypothesis of Lemma \ref{monotone} and concludes the proof of Theorem 
\ref{existence}.

\hfill $\Box $

Turning back to the Coulomb system problem, some remarks are now in order.

\begin{remark}
Recalling $v(t,x)=u_{x}(t,x)$ and denoting $\lambda ^{\ast
}=\lim\limits_{n\rightarrow \infty }\lambda ^{n}$ the charge activity at the
fixed point, we have from (\ref{utx}) 
\begin{equation*}
\psi (0)=-i\sum_{q\in \mathbb{Z}}q\,\lambda ^{\ast }(q)\left/ \sum_{q\in 
\mathbb{Z}}\,\lambda ^{\ast }(q)\right. =0
\end{equation*}
and 
\begin{equation*}
\psi ^{\prime }(0)=\sum_{q\in \mathbb{Z}}q^{2}\,\lambda ^{\ast }(q)\left/
\sum_{q\in \mathbb{Z}}\,\lambda ^{\ast }(q)\right. \geq 0\,.
\end{equation*}
These boundary conditions select $\psi _{j}^{+}$, $j=1,\ldots ,k$, as being
the only physically meaningful stationary solutions and implies $\phi
^{+}(x)=\displaystyle\int_{0}^{x}\psi ^{+}(y)\,dy\geq 0$ on $\left( -\pi
,\pi \right) $.
\end{remark}

\begin{remark}
The value $\alpha =2$ is a bifurcation point as one can see by linearizing
(\ref{stationary}) about $\psi \equiv 0$. The linear operator $L[0]=A$ given 
by (\ref{A}) in the subspace of odd $2\pi $--periodic functions has
eigenvalues and associate eigenfunctions given by (\ref{spectrum}). Hence,
if $\alpha >2$, the eigenvalues are all positive and $\psi \equiv 0$ is
locally stable. When $\alpha <2$ (but close to $2$) a single eigenvalue
becomes negative and one can apply the Crandall--Rabinowitz bifurcation
theory \cite{C} to locally describe the stable solution which bifurcates
from the trivial one. Note that Crandall--Rabinowitz theory can also be
applied in the neighborhood of $\alpha _{j}=2/j^{2}$, $j>1$, in the
orthogonal complement of the span $\left\{ \pi ^{-1/2}\sin
mx,\,m=1,...,j-1\right\} $ corresponding to the odd functions with
fundamental period $T=2\pi /j$. These points were referred to in the
introduction as a sequence of instability thresholds.

In Theorem \ref{existence} we have given a global characterization of the
non--trivial stationary solutions.
\end{remark}

\begin{remark}
\label{sinegordon} In the sine--Gordon representation, the effective
potential $\phi (x)=\int_{0}^{x}\psi (y)\,dy=x^{2}/\left( 2\alpha \right) $
at $\gamma _{\alpha ^{-1}}$\ corresponds the Debye--H\"{u}ckel regime with
Debye length $\alpha $. Although this regime is not reached for all $\beta >0
$, it gets closed quite fast as $\beta =4\pi \alpha $ approaches $0$.
Numerical calculation is shown in Figure 2. Note that at $\alpha =1$ ($\beta
=4\pi $), $\widehat{w}_{1}$ cannot be distinguished from $\alpha ^{-1}$
(numerical error is in the sixth decimal order).
\end{remark}


\begin{figure}[hbtp]
\begin{center}
\epsfig{file=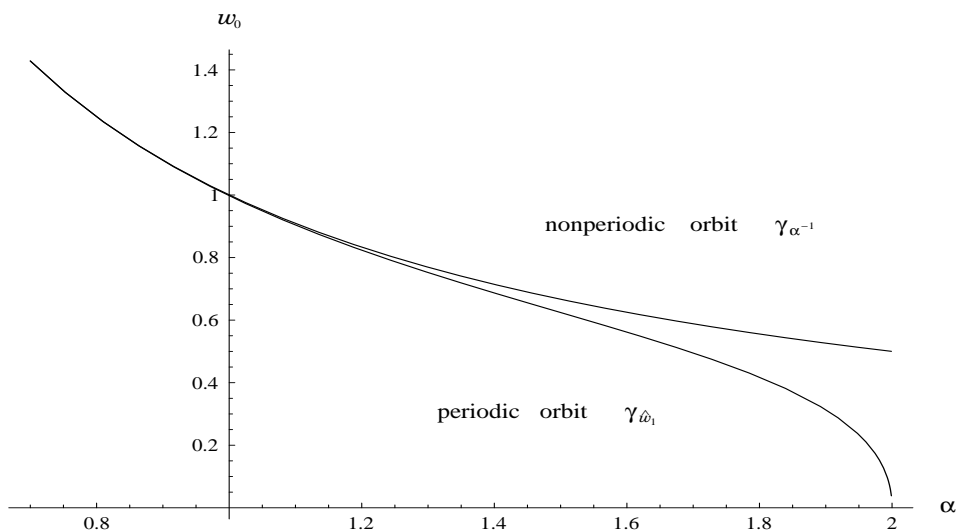, height=7cm, width=13.5cm, clip=}
\end{center}
\caption{Comparison between the initial value function for the periodic orbit
of period $2\pi $, ${\widehat w}_{1}={\widehat
w}_{1} (\alpha )$,  and for the nonperiodic
Debye--H\"{u}ckel orbit, ${\widehat w}_{\rm DH} (\alpha ) = \alpha ^{-1}$.}
\label{debey}
\end{figure}


\begin{remark}
The derivative of (\ref{T}) with respect to $w_{0}$, computed from equation
(\ref{h2}),  
\begin{equation*}
\frac{\partial T}{\partial w_{0}}=\frac{2\alpha w_{0}}{1-\alpha w_{0}}\int 
\mathrm{sign}\left( w\right) \frac{2\left( 1-\alpha w\right) p^{2}/\alpha }{
\left( 1+2\left( 1-\alpha w\right) p^{2}/\alpha \right) }dp\,,
\end{equation*}
indicates that an estimate from below can be very delicate to obtain. Note $
\mathrm{sign}\left( w\right) $ changes along the orbit $\gamma _{w_{0}}$.
This shows how amusing Chicone's monotonicity result is for the problem at
hand.
\end{remark}


\section{Stability \label{Stability}}

\setcounter{equation}{0} \setcounter{theorem}{0}

Let $z(t;z_{0})$ denote the solution of \ the initial value problem
(\ref{ode}) -- (\ref{A}). It follows from the analysis in Section \ref{EU}
that  
\begin{equation}
S(t)z_{0}=z(t;z_{0})  \label{dyn}
\end{equation}
defines a dynamical system on a closed subset $\mathcal{V}\subset D\left(
A\right) $ of $\mathcal{B}^{1/2}$ with the topology induced by the graph
norm $\left\| \cdot \right\| _{1/2}$. Note that $z(t;z_{0})$ is continuous
in both $t$ and $z_{0}$ with $z(0;z_{0})=z_{0}$ and satisfies the
(nonlinear) semi--group property $S(t+\tau )z_{0}=z(t;z(\tau
;z_{0}))=S(t)S(\tau )z_{0}$.

This section is devoted to the stability analysis of the equilibrium
solutions described in Section \ref{SS}. By local stability it is meant that 
$z(t;z_{0})$ is uniformly continuous in $\mathcal{V}$ for all $t\geq 0$:
given $\varepsilon >0$, $\left\| z(t;z_{0})-z(t;z_{1})\right\|
_{1/2}<\varepsilon $ for all $t\geq 0$ and $z_{1}\in \mathcal{V}$ such $
\left\| z_{1}-z_{0}\right\| _{1/2}<\delta $ for some $\delta =\delta
(\varepsilon )>0$. It is uniformly asymptotically stable if, in addition, $
\lim\limits_{t\rightarrow \infty }\left\| z(t;z_{0})-z(t;z_{1})\right\|
_{1/2}=0$.

The Liapunov (global) stability analysis as developed by LaSalle and applied
to semilinear parabolic differential equations by Chafee and Infante \cite{CI}
(see also \cite{H}) will also be discussed and extended in this section. 

Let us begin with the local analysis.

\begin{theorem}[Local Stability]
\label{lstb} There exists a neighborhood $\mathcal{U}\subset \mathcal{B}
^{1/2}$ of the origin such that, if $\alpha >2$ and $z_{0}$ in $\mathcal{U}$
, then $\psi _{0}\equiv 0$ is stable, i.e., $\lim\limits_{t\rightarrow
\infty }\left\| z(t;z_{0})\right\| _{1/2}=0$. If $\alpha <2$ is such that $
2/\left( k+1\right) ^{2}\leq \alpha <2/k^{2}$ holds, among all equilibrium
solutions of (\ref{stationary}), $\psi _{0},\psi _{j}^{\pm }$, $j=1,\ldots ,k
$, $\psi _{1}^{\pm }$ are the only asymptotically stables. So, there exists $
\rho >0$ such that if $\left\| z_{0}-\psi \right\| _{1/2}\leq \rho $, then $
\lim\limits_{t\rightarrow \infty }\left\| z(t;z_{0})-\psi \right\| _{1/2}=0$
for $\psi =\psi _{1}^{\pm }$ and, for any sequence $\left\{ z_{n}\right\}
_{n\geq 1}$ with $\lim\limits_{n\rightarrow \infty }\left\| z_{n}-\psi
\right\| =0$, we have $\sup\limits_{t>0}\left\| z(t;z_{n})-\psi \right\|
_{1/2}\geq \varepsilon >0$ for all $n$ and $\psi =\psi _{j}^{\pm }$, $j\neq 1
$.
\end{theorem}

It is convenient to consider the equation

\begin{equation}
\frac{d\zeta }{dt}+L\zeta =F\left( \zeta \right)  \label{ode1}
\end{equation}
for $\zeta =z-\psi $ where $\psi $ is a solution of (\ref{stationary}). Here 
\begin{equation}
L\zeta =L\left[ \psi \right] \zeta =-\alpha \zeta ^{\prime \prime }+2\alpha
\psi \zeta ^{\prime }-2\left( 1-\alpha \psi ^{\prime }\right) \zeta
\label{L}
\end{equation}
is the linearization of the differential operator (\ref{dpde}) around $\psi $
and $F$ is as in (\ref{A}). Note $L=A$ and (\ref{ode1}) reduces to (\ref{ode})
if $\psi =\psi _{0}=0$. 

\noindent \textbf{Proof.} The proof of the Theorem \ref{lstb} follows from the
next two theorems. 

\begin{theorem}
\label{asym} If the spectrum $\sigma (L)$ of (\ref{L}) lies in $\left\{
\lambda \in \mathbb{R}:\lambda \geq c\right\} $ for some $c>0$, then $\zeta
=0$ is the unique uniformly asymptotically stable solution of (\ref{ode1}).
On the other hand, if $\sigma (L)\cap \left\{ \lambda \in \mathbb{R}:\lambda
<0\right\} \neq \emptyset $, then $\zeta =0$ is unstable.
\end{theorem}

\begin{theorem}
\label{positive} Let $L=L[\psi ]$\ be given by (\ref{L}). Then $\sigma (L)>0$
whenever $\psi =\psi _{0}$ and $\alpha >2$ or $\psi =\psi _{1}^{\pm }$ and $
\alpha <2$. If $\alpha $ is such that\ $2/\left( k+1\right) ^{2}\leq \alpha
<2/k^{2}$ holds for some $k\in \mathbb{N}_{+}$, then $\sigma (L)\cap \left\{
\lambda \in \mathbb{R}:\lambda <0\right\} \neq \emptyset $ for $\psi =\psi
_{0}$ and $\psi =\psi _{j}^{\pm },\;j=2,\ldots ,k$.
\end{theorem}

\noindent \textbf{Proof of Theorem \ref{asym}.} We shall prove only the
first part of Theorem \ref{asym} and refer to Theorem 5.1.3 of Henry's book 
\cite{H} for the instability part.

It follows from (\ref{integral}), (\ref{b1}), (\ref{lipsch}) and the
hypothesis on $\sigma (L)$ that 
\begin{equation}
\left\| \zeta (t)\right\| _{1/2}\leq C_{1/2}\,e^{-ct}\left\| \zeta
_{0}\right\| _{1/2}+\xi \int_{0}^{t}\,\left( t-s\right)
^{-1/2}\,e^{-c(t-s)}\left\| \,\zeta (s)\right\| _{1/2}^{2}\,\,ds\,,
\label{zt}
\end{equation}
with $c>0$, $C_{1/2}=1/\sqrt{2e}$ and $\xi =2\sqrt{2\pi }\alpha $. \ 

Let us assume that $\left\| \zeta (s)\right\| _{1/2}\leq \rho $ on a
interval $\left( 0,t\right) $ for some $\rho $ satisfying 
\begin{equation}
\xi \int_{0}^{\infty }t^{-1/2}\,e^{-ct}\,dt=\xi \sqrt{\frac{\pi }{c}}<\frac{
1 }{2\rho },  \label{rho}
\end{equation}
i. e., $\rho <\dfrac{1}{4\pi \alpha }\sqrt{\dfrac{c}{2}}$. If $\left\| \zeta
_{0}\right\| _{1/2}\leq \rho \sqrt{\dfrac{e}{2}}$, then equation (\ref{zt})
can be bounded as 
\begin{equation}
\left\| \zeta (t)\right\| _{1/2}\leq \frac{\rho }{2}+\rho ^{2}\xi
\int_{0}^{t}\,\left( t-s\right) ^{-1/2}\,e^{-c(t-s)}<\rho \,  \label{zt1}
\end{equation}
and this implies the existence of a unique solution of (\ref{ode1}) with $
\left\| \zeta (t)\right\| _{1/2}\leq \rho $ for all $t>0$. Note that $
\left\| \zeta _{0}\right\| _{1/2}<\rho $ and if $t_{1}$ is the maximum value
under which $\left\| \zeta (t)\right\| _{1/2}<\rho $ for all $0<t<t_{1}$,
then either $\left\| \zeta (t_{1})\right\| _{1/2}=\rho $ or $t_{1}=\infty $.
But the first case is impossible by (\ref{zt1}).

Going back to (\ref{zt}), using $\left\| \,\zeta (s)\right\| _{1/2}<\rho $
and a slightly modification of Gronwall inequality (\ref{gronwall}) with $
E_{1/2}(t)=$ $\displaystyle\sum\limits_{n=0}^{\infty }\left( \rho \xi \sqrt{
\pi }t^{1/2}\right) ^{n}\left/ \Gamma (n/2+1)\right. $, we have 
\begin{eqnarray*}
\left\| \zeta (t)\right\| _{1/2} &\leq &C_{1/2}\left\| \zeta _{0}\right\|
_{1/2}E_{1/2}(t)\,\,e^{-ct} \\
&\leq &C_{1/2}\left\| \zeta _{0}\right\| _{1/2}\left( 1+\rho \xi
t^{1/2}\right) e^{-\left( c-\rho ^{2}\xi ^{2}\pi \right) t} \\
&\leq &\frac{1}{2e}\left\| \zeta _{0}\right\| _{1/2}\left( 1+\frac{1}{2} 
\sqrt{\frac{ct}{\pi }}\right) \,e^{-3ct/4}\,,
\end{eqnarray*}
in view of (\ref{rho}). This proves the stability statement of Theorem
\ref{asym}, since (\ref{ode1}) defines a dynamical system in a closed subset
$\mathcal{V}_{\rho }=\left\{ \zeta \in \mathcal{B}^{1/2}:\left\| \zeta
\right\| _{1/2}\leq \rho \right\} $ with $\lim\limits_{t\rightarrow \infty
}\left\| \zeta (t)\right\| _{1/2}=0$ if $\left\| \zeta _{0}\right\|
_{1/2}=\left\| z_{0}-\psi \right\| _{1/2}\leq \rho \sqrt{\dfrac{e}{2}}$.

\hfill $\Box $

\begin{remark}
One can actually show that if $c=\inf_{\lambda }\sigma (L)$ then $\zeta
(t;\zeta _{0})=z(t;z_{0})-\psi $ decays exponentially fast to $0$ as 
\begin{equation*}
\zeta (t;\zeta _{0})=\kappa (\zeta _{0})\,e^{-ct}+\varepsilon (t;\zeta _{0})
\end{equation*}
where $\left\| \varepsilon (t;\zeta _{0})\right\| _{1/2}\leq C\left\| \zeta
_{0}\right\| _{1/2}e^{-c^{\prime }t}$ with $0<c<c^{\prime }$ and $\kappa :
\mathcal{V}_{\rho }\longrightarrow \mathcal{N}\left( L-cI\right) $ is
continuous and such that $\kappa (0)=0$, where $\mathcal{N}\left(
L-cI\right) $ is the one--dimensional span of the eigenfunction of $L$
associated to $c$.
\end{remark}

\noindent \textbf{Proof of Theorem \ref{positive}. }Since $L[\psi _{0}]=A$,
Theorem \ref{positive} for $\psi =\psi _{0}$ with $\alpha \geq 0\ $follows
from the spectral computation in (\ref{spectrum}).

Now, let $\psi $ be a nontrivial solution of the equilibrium equation
(\ref{stationary}) and note that $\psi (0)=\psi (\pi )=0$ by parity. According
to Theorem \ref{asym}, $\psi $ is asymptotically stable if $\sigma (L)>0$ and
unstable if $\sigma (L)\cap \left\{ \lambda <0\right\} \neq \emptyset $.

Let $\varphi $ be the solution of 
\begin{equation}
L[\psi ]\varphi =0  \label{Lphi}
\end{equation}
in the domain $0<x<\pi $ satisfying 
\begin{equation}
\varphi (0)=0\qquad \mathrm{and}\qquad \varphi ^{\prime }\left( 0\right) =1.
\label{phiphi}
\end{equation}
As in \cite{H}, we shall use the comparison theorem to establish that $\psi $
is asymptotically stable if $\varphi (x)>0$ on $0<x\leq \pi $ and unstable
if $\varphi (x)<0$ somewhere in $0<x<\pi $. \ 

To apply the comparison theorem and complete the proof of Theorem
\ref{positive}, let  
\begin{equation}
p(x):=e^{-2\int_{0}^{x}\psi (y)\,dy}  \label{p(x)}
\end{equation}
be the weight which makes $L$ a self--adjoint operator: 
\begin{equation}
p\,L[\psi ]\zeta =-\alpha \left( p\,\zeta ^{\prime }\right) ^{\prime
}-2p\left( 1-\alpha \psi ^{\prime }\right) \zeta \,.  \label{pL}
\end{equation}
Note that $\left( L\zeta ,\eta \right) _{p}=\left( \zeta ,L\eta \right) _{p}$
for any odd periodic functions $\zeta $ and $\eta $ of period $2\pi $ were $
\left( f,g\right) _{p}:=\displaystyle\int_{-\pi }^{\pi}f(x)\,g(x)\,p(x)\,dx$.

\begin{theorem}[Comparison]
\label{comp} Suppose $\zeta _{1}$ and $\zeta _{2}$ are two real solutions on
the domain $\left( 0,\pi \right) $ of 
\begin{equation*}
p\,L[\psi ]\zeta =f_{i}\,,\qquad i=1,2\,,
\end{equation*}
respectively, with $\zeta _{1}(0)=\zeta _{1}(\pi )=0$, $\zeta _{1}^{\prime
}(0)>0$ and $\zeta _{2}(0)=0$, $\zeta _{2}^{\prime }(0)>0$. If $\zeta _{1}>0$
and $f_{i}=f_{i}(\zeta ;x)$ is such that 
\begin{equation}
f_{2}>f_{1}  \label{f2f1}
\end{equation}
on $\left( 0,\pi \right) $, then $\zeta _{2}$ must vanish at some point of
this domain.
\end{theorem}

\noindent \textbf{Proof. }Let assume that $\zeta _{2}>0$ on $\left( 0,\pi
\right) $. Then, from (\ref{pL}) and the hypotheses of Theorem \ref{comp},
we have 
\begin{eqnarray*}
2\int_{0}^{\pi }\,\left( f_{2}-f_{1}\right) \,dx &=&\left( \zeta _{1},L\zeta
_{2}\right) _{p}-\left( L\zeta _{1},\zeta _{2}\right) _{p} \\
&=&2\alpha \int_{0}^{\pi }\left[ \left( p\,\zeta _{1}^{\prime }\right)
^{\prime }\zeta _{2}-\zeta _{1}\left( p\,\zeta _{2}^{\prime }\right)
^{\prime }\right] \,dx \\
&=&2\alpha \int_{0}^{\pi }\left[ p\,\left( \zeta _{1}^{\prime }\zeta
_{2}-\zeta _{1}\,\zeta _{2}^{\prime }\right) \right] ^{\prime }\,dx
\end{eqnarray*}
which, in view of the boundary conditions and (\ref{f2f1}), implies a
contradiction 
\begin{equation*}
p(\pi )\,\zeta _{1}^{\prime }(\pi )\,\zeta _{2}(\pi )>0\,.
\end{equation*}
Note that $\zeta _{1}^{\prime }(\pi )<0$ since $\zeta _{1}>0$ on $\left(
0,\pi \right) $ and $\zeta _{1}(\pi )=0$. So, there must exist $\overline{x}
\in \left( 0,\pi \right) $ such that $\zeta _{2}(\overline{x})=0$.

\hfill $\Box $

If we consider the eigenvalue equation 
\begin{equation}
L[\psi ]\theta =\lambda \theta  \label{L-l}
\end{equation}
on $\left( 0,\pi \right) $ for the smallest eigenvalue $\lambda $ in the
space of odd periodic function, $\theta $ satisfies the conditions of $\zeta
_{1}$ in Theorem \ref{comp} with $f_{1}=\lambda p\zeta $. Note the
eigenfunction associated to the smallest eigenvalue may be chosen to be
positive in the domain $\left( 0,\pi \right) $.

Applying Theorem \ref{comp} for (\ref{Lphi}) and (\ref{L-l}) we arrive to
the following stability criterium:

\begin{criterium}
\label{critm}The smallest eigenvalue $\lambda $ of $L[\psi ]$ is positive if 
$\varphi >0$ on $\left( 0,\pi \right) $ and negative if there exist $
\overline{x}\in \left( 0,\pi \right) $ such that $\varphi (\overline{x})=0$,
where $\varphi $ is the solution of equations (\ref{Lphi}) and
(\ref{phiphi}). 
\end{criterium}

\smallskip

Now,\ for a given non--trivial stationary solution $\psi $ let 
\begin{equation}
\chi =c\left( -\alpha \psi ^{\prime \prime }+4\psi \right) \,,  \label{chi}
\end{equation}
where $c>0$ is chosen so that $\chi ^{\prime }(0)=1$. It follows from the
equation $-\alpha \psi ^{\prime \prime }=2\left( 1-\alpha \psi ^{\prime
}\right) \psi \,$\ (see (\ref{stationary})), that 
\begin{equation*}
\chi (0)=0\qquad \qquad \mathrm{and}\qquad \qquad \chi >0
\end{equation*}
whenever $\psi >0$ (recall $\psi (0)=0$ and $1-\alpha \psi ^{\prime }>0$ for
all closed orbits). Moreover, we have

\begin{proposition}
\label{equal} 
\begin{equation}
L[\psi ]\chi =8c\alpha ^{2}\psi \left( \psi ^{\prime }\right) ^{2}>0
\label{Lchi}
\end{equation}
on the same domain $\left( 0,\overline{x}\right) $ that $\psi >0$.
\end{proposition}

\noindent \textbf{Proof. }Differentiating (\ref{stationary}) twice, 
\begin{equation*}
-\alpha \left( \psi ^{\prime \prime }\right) ^{\prime \prime }=-2\alpha \psi
\left( \psi ^{\prime \prime }\right) ^{\prime }+2\left( 1-3\alpha \psi
^{\prime }\right) \psi ^{\prime \prime }\,,
\end{equation*}
and using (\ref{stationary}) again, gives 
\begin{eqnarray*}
L[\psi ]\psi ^{\prime \prime } &=&-\alpha \left( \psi ^{\prime \prime
}\right) ^{\prime \prime }+2\alpha \psi \left( \psi ^{\prime \prime }\right)
^{\prime }-2\left( 1-\alpha \psi ^{\prime }\right) \psi ^{\prime \prime } \\
&=&-4\alpha \psi ^{\prime }\psi ^{\prime \prime } \\
&=&8\left( 1-\alpha \psi ^{\prime }\right) \psi \,\psi ^{\prime }
\end{eqnarray*}
In addition, we have 
\begin{eqnarray*}
L[\psi ]\psi &=&-\alpha \psi ^{\prime \prime }+2\alpha \psi \psi ^{\prime
}-2\left( 1-\alpha \psi ^{\prime }\right) \psi \\
&=&2\alpha \psi \psi ^{\prime }
\end{eqnarray*}
which combined with the above equation, gives the equality in Proposition 
\ref{equal}.

\hfill $\Box $

\noindent \textbf{Completion of the proof of Theorem \ref{positive}.} We are
in position to prove Theorem \ref{positive} for non--trivial equilibrium
solutions. Let $\chi $ be given by (\ref{chi}) with $\psi =\psi _{1}^{+}$.
Then $\chi >0$ on $\left( 0,\pi \right) $ and Theorem \ref{comp} can be used
to compare equation (\ref{Lchi}) with (\ref{Lphi}). This yields $\varphi
>\chi \geq 0$ on $(0,\pi ]$ which implies the stability of $\psi _{1}^{+}$
by Criterium \ref{critm}.

For instability, we observe that $\psi ^{\prime }$ satisfies 
\begin{eqnarray*}
L[\psi ]\psi ^{\prime } &=&-\alpha \psi ^{\prime \prime \prime }+2\alpha
\psi \psi ^{\prime \prime }-2\left( 1-\alpha \psi ^{\prime }\right) \psi
^{\prime } \\
&=&\left( -\alpha \psi ^{\prime \prime }+2\alpha \psi \psi ^{\prime }-2\psi
\right) ^{\prime }=0\,,
\end{eqnarray*}
in view of equation (\ref{stationary}). Recall that $\psi =\psi _{j}^{+}$
with $j\geq 2$, has fundamental period $2\pi /j$ and satisfies $\psi (\pi
/j)=\psi ^{\prime \prime }(\pi /j)=0$ by the odd parity and equation
(\ref{stationary}). Since $\psi ^{\prime }(0)>0$, this implies $\psi <0$ on $
\left( \pi /j,2\pi /j\right) $ and the minimum of $\psi $ is attained at $
\underline{x}=\dfrac{3\pi }{2j}$. Since $\psi ^{\prime }$ and $\varphi $
satisfies the same self--adjoint equation $pL[\psi ]\zeta =0$, their
Wronskian 
\begin{eqnarray*}
W\left( \varphi ,\psi ^{\prime };x\right) &=&\left| 
\begin{array}{cc}
\varphi & \psi ^{\prime } \\ 
-\alpha p\varphi ^{\prime } & -\alpha p\psi ^{\prime \prime }
\end{array}
\right| \\
&=&\alpha p\left( \varphi ^{\prime }\psi ^{\prime }-\varphi \psi ^{\prime
\prime }\right) =\alpha \psi ^{\prime }(0)>0
\end{eqnarray*}
is a non--vanishing constant (recall $p(0)=1$, $\varphi (0)=0$ and $\left(
\psi _{j}^{+}\right) ^{\prime }(0)>0$). As a consequence 
\begin{equation*}
W\left( \varphi ,\psi ^{\prime };\pi /j\right) =-\alpha p(\underline{x}
)\varphi \left( \underline{x}\right) \psi ^{\prime \prime }\left( \underline{
x}\right) >0
\end{equation*}
implies $\varphi \left( \underline{x}\right) <0$ because $\psi ^{\prime
\prime }\left( \underline{x}\right) >0$. It thus follows from the stability
criterium that $\psi _{j}^{+},\,j=2,\ldots ,k$, are unstable since $
\underline{x}\in \left( 0,\pi \right) $ provided $j\geq 2$ and there exist $
\overline{x}\in \left( 0,\pi \right) $, $\overline{x}<\underline{x}$, such
that $\varphi (\overline{x})=0$.

By a slight modification of these arguments, one may conclude the stability
of $\psi _{1}^{-}$and instability of $\psi _{j}^{-},\,j=2,\ldots ,k$, as
well. This concludes the proof of Theorem \ref{positive} and, consequently,
the proof of Theorem \ref{lstb}.

\hfill $\Box $

\medskip

Now, we turn to the Liapunov stability analysis with a proof of global
stability of the trivial solution $\phi _{0}\equiv 0$.

Let $V$ be a real--valued functional on the subspace of absolutely
continuous function of $D(A)$ given by 
\begin{equation}
V\left( v\right) =\int_{-\pi }^{\pi }\left\{ \left( \alpha ^{-1}-\,v^{\prime
}\right) \ln \left( 1-\alpha v^{\prime }\right) +v^{\prime }-v^{2}\right\}
\,dx  \label{V}
\end{equation}
and notice that $V(0)=0$ and $V\left( \eta \right) =W(\eta )+o\left( \left\|
\eta \right\| ^{2}\right) $, as $\left\| \eta \right\| \rightarrow 0$, where 
\begin{equation*}
W\left( v\right) =\frac{1}{2}\int_{-\pi }^{\pi }\left( \alpha \,v^{\prime
2}-2v^{2}\,\right) \,dx\,,
\end{equation*}
by Taylor expanding $g(w)=\left( \alpha ^{-1}-w\right) \ln \left( 1-\alpha
w\right) -w$ around $w=0$. Observe that $W\left( v\right) =\left( 1/2\right)
\left\| v\right\| _{1/2}^{2}$ if $\alpha >2$ and since $g(w)-(\alpha
/2)w^{2}\geq 0$ if $\alpha w<1$, $V\left( v\right) \leq W\left( v\right) $
holds on the space 
\begin{equation*}
\mathcal{V}=\left\{ v\in H_{\mathrm{o,p}}^{1}\cap H_{\mathrm{o,p}
}^{2}:\alpha v^{\prime }<1\right\} \,\,, 
\end{equation*}
of odd, positive and $2\pi $--periodic functions with distributional
derivative up to second order.

A Liapunov function $V$ of a dynamical system $\left\{ S(t),t\geq 0\right\} $
satisfies 
\begin{equation}
\overset{\cdot }{V}(v)=\overline{\lim\limits_{t\downarrow 0}}\frac{1}{t} 
\left( V\left( S(t)v\right) -V(v)\right) \leq 0  \label{Vdot} 
\end{equation}
for all $v\in \mathcal{V}$. We now show that (\ref{Vdot}) holds if $S(t)$ is
given by equation (\ref{dpde}). More precisely,

\begin{proposition}
Let $S(t)v_{0}=v(t;v_{0})$ be the dynamical system in $\mathcal{V}$ given by
(\ref{dyn}). Then, the pair of functions 
\begin{equation}
\rho (w)=\frac{1}{1-\alpha w}  \label{rhow}
\end{equation}
and 
\begin{equation}
\Phi (p,w)=\left( \alpha ^{-1}-\,w\right) \ln \left( 1-\alpha w\right)
+w-p^{2}  \label{Ppw}
\end{equation}
generate the Liapunov function given by (\ref{V}): 
\begin{equation}
V(v)=\int_{0}^{\pi }\Phi (v,v_{x})\,dx\qquad \mathrm{with}\qquad \overset{
\cdot }{V}(v)=\int_{0}^{\pi }\rho (v_{x})v_{t}^{2}\,dx\,.  \label{VVv}
\end{equation}
\end{proposition}

\noindent \textbf{Proof.} Note that, from the parity of $v$ the integral in
(\ref{V}) can be made over $[0,\pi ]$. By the calculus of variations and 
equations (\ref{rhow}) and (\ref{Ppw}) we have 
\begin{eqnarray}
\dot{V}\left( v\left( t,\cdot \right) \right) &=&-\int_{0}^{\pi }\left( 
\frac{d}{dx}\frac{\partial \Phi }{\partial v_{x}}-\frac{\partial \Phi }{
\partial v}\right) v_{t}\,dx+\left. \frac{\partial \Phi }{\partial v_{x}}
v_{t}\right| _{0}^{\pi }  \notag \\
&=&-\int_{0}^{\pi }\left( -\frac{d}{dx}\ln (1-\alpha v_{x})+2v\right)
v_{t}\,dx  \notag \\
&=&-\int_{0}^{\pi }\rho (v_{x})\,\left( \alpha v_{xx}-2\alpha
vv_{x}+2v\right) v_{t}\,dx\,,  \label{Vv}
\end{eqnarray}
where $v_{t}(t,0)=v_{t}(t,\pi )=0$, $t\geq 0$, in view of the boundary
conditions on $\mathcal{V}$. Since $\rho (w)\geq 0$ for $\alpha w<1$, this
with (\ref{dpde}) concludes the proof of Proposition.

\hfill $\Box $

\begin{remark}
We have used the construction method based in the Euler--Lagrange equation
to find this Liapunov function (see e.g. Chap. 2 of Zelenyak, Lavrentiev and
Vishnevskii \cite{ZLV}). A sufficient condition for (\ref{VVv}) hold leads
to a first order partial differential equation for $\rho $ 
\begin{equation*}
w\rho _{p}-\frac{2}{\alpha }(1-\alpha w)p\rho _{w}=-2p\rho 
\end{equation*}
whose characteristics are given by the orbits $\gamma _{w_{0}}$ described in
Section \ref{SS} in the study of the equilibrium solutions of (\ref{dpde}).
Note that equation (\ref{Ppw}) is the Lagrangian associated with the
Hamiltonian (\ref{hamiltonian}) (with $q$ defined by (\ref{phi})). Due to
the requirement $\alpha w<1$, our particular solution takes into account
only the closed orbits. There may be other suitable choices which includes
all orbits.
\end{remark}

The proof of global stability of $\phi _{0}$ requires that a subspace of $
\mathcal{V}$ be invariant under the flow equation (\ref{dpde}). This is
shown in the following by using the maximum principle.

\begin{theorem}
\label{cone} If $v(t,x)$ is a classical solution of equation (\ref{dpde})
with initial condition $v(0,x)=v_{0}(x)\in \mathcal{V}$, then $\alpha
v_{x}(t,x)<1$ and 
\begin{equation}
\alpha ^{-1}\left( x-\pi \right) <v(t,x)<\alpha ^{-1}x\,,  \label{tud}
\end{equation}
hold for all $t\geq 0$ and $0\leq x\leq \pi $.
\end{theorem}

\noindent \textbf{Proof.} Denoting 
\begin{equation}
L[v]:=F(v_{xx},v_{x},v)-v_{t},  \label{Fut}
\end{equation}
where $F(a_{1},a_{2},a_{3})=\alpha \left( a_{1}-2a_{2}\,a_{3}\right)
+2a_{3}\ $is a continuous and differentiable function of its variables, the
differential equation (\ref{dpde}) can be written as 
\begin{equation}
L[v]=0.  \label{Lv}
\end{equation}

For $v$ satisfying (\ref{Lv}) with $v(t,0)=v(t,\pi )=0\,$, $0\leq t\leq \tau 
$, and initial data $v(0,\cdot )=v_{0}$, let us suppose $z=z(t,x)$ and $
Z=Z(t,x)$ are such that 
\begin{equation}
L[Z]\leq 0\leq L[z]  \label{LZLz}
\end{equation}
for all $\left( t,x\right) $ in $D=\left( 0,\tau \right) \times \left( 0,\pi
\right) $ with $z(t,y)\leq 0\leq Z(t,y)$, $y=0,\pi $ and $0\leq t\leq \tau $
, and 
\begin{equation*}
z(0,x)\leq v_{0}(x)\leq Z(0,x)\,,
\end{equation*}
for $0\leq x\leq \pi $. Then, by the maximum principle (see \cite{PW},
Theorem $12$ in Chap. $3$), 
\begin{equation}
z(t,x)\leq v(t,x)\leq Z(t,x)\,  \label{zuZ}
\end{equation}
in $\overline{D}=[0,\pi ]\times \lbrack 0,\tau ]$.

The lower limit function $z$ is given by 
\begin{equation}
z(x)=\theta \left( x-\pi \right) \,,  \label{z}
\end{equation}
with $\theta \geq 0$. From (\ref{Lv}), 
\begin{equation*}
L[z]=2\theta (\alpha \theta -1)(\pi -x)\}\,,
\end{equation*}
is always positive provided $\alpha \theta \geq 1$.

Analogously, the upper limit function $Z$ is given by 
\begin{equation}
Z(x)=\delta x\,,  \label{Z}
\end{equation}
from which 
\begin{equation*}
L[Z]=-2\delta (\alpha \delta -1)x\,
\end{equation*}
is always negative provided $\alpha \delta \geq 1$.

Since (\ref{LZLz}) holds uniformly in $\tau $, equation (\ref{zuZ}) holds
for all $(t,x)$ in $\mathbb{R}_{+}\times \lbrack 0,\pi ]$. Note that $v$
remains bounded irrespective of $\alpha v_{x}<1$. However, if this condition
holds for $t=0$, it remains for all $t>0$. To see this, observe from the
equation $v_{t}=v_{xx}+(1-\alpha v_{x})v$ with $v_{xx}=0$, that the rate by
which $|v|$ increases tends to zero when the inequality saturates. The
inclusion of the Laplacian only smooths $v$ and prevents, even more, $v_{x}$
to increase beyond the threshold. The same argument justify the strict
inequality (\ref{tud}).

This concludes proof of Theorem \ref{cone}.

\hfill $\Box $

We pause to discuss some properties of the classical solutions of equations
(\ref{pde}) and (\ref{utildeq}). Recall that $\widetilde{u} (t,x)=
\displaystyle\int_{0}^{x}v(t,y)\,dy$ with $v$ satisfying (\ref{dpde}).

\begin{remark}
Note that the cone $\mathcal{C}=\left\{ u\in H_{\mathrm{e,p}}^{1}\cap H_{
\mathrm{e,p}}^{2}:u\geq 0,\,\alpha u_{xx}<1\right\} $ is invariant under the
unnormalized evolution (\ref{pde}). For this, let 
\begin{equation*}
M[u]:=\alpha (u_{xx}-u_{x}^{2})+2u.
\end{equation*}
If $u(t,x)$ is a classical solution of (\ref{pde}) with initial value $
u_{0}\in \mathcal{C}$, since $M[u]=0$ for $u\equiv 0$, we have by Theorem $7$
in Chap. $3$ of \cite{PW} (see also Remark (ii) after this) that $u(t,x)\geq
0$ for all $t>0$. This, however, does not imply that $\widetilde{u}(t,x)$
remains positive (recall (\ref{uutilde})). A proof of this assertion goes as
follows.

Theorem \ref{cone} implies $\widetilde{u}(t,x)$ remains bounded, and $
\widetilde{u}_{xx}(t,0)$ bounded from above, if $\widetilde{u}$ satisfies
(\ref{utildeq}) with initial condition $u_{0}$ satisfying $\alpha 
^{-1}[(x-\pi )^{2}/2-\pi ^{2}/2]<u_{0}<\alpha ^{-1}x^{2}/2$ (by integrating
(\ref{tud})). The comparison principle applied directly to equation
(\ref{utildeq}) leads to (\ref{LZLz}) with $L$ replaced by $M$ and $0$
replaced by $\alpha \widetilde{u}_{xx}(t,0)$. An upper and lower solutions,
$z$ and $Z $, can be obtained from the solution of the equilibrium initial value
problem (\ref{syst}): 
\begin{equation*}
\Phi ^{\pm }(\alpha ,w_{0};x)=\int_{0}^{x}\Psi ^{\pm }(\alpha
,w_{0};y)\,dy\,,
\end{equation*}
where $\Psi ^{\pm }(\alpha ,w_{0};x)$ is the $p$--component of the closed
orbit $\gamma _{\pm w_{0}}$ starting at $(\pm w_{0},0)$. Note that $\Phi
^{\pm }$ is an even periodic function of period $T=T(\alpha ,w_{0})$ given
by a monotonically increasing function of both $w_{0}$ and $\alpha $ with $
T\rightarrow \infty $ as $\alpha w_{0}\uparrow 1$ for $\Phi ^{+}$ and as $
w_{0}\rightarrow \infty $ for $\Phi ^{-}$ (see proof of Theorem
\ref{existence}\ for details). $\Phi ^{+(-)}$ is also a monotone increasing 
(decreasing) function of $x$ in $[0,T/2]$ and satisfies 
\begin{equation*}
M[\Phi ^{\pm }]=\pm \alpha w_{0}\,,
\end{equation*}
with $\Phi ^{\pm }(0)=\Phi ^{\pm \prime }(0)=\Phi ^{\pm \prime }(T/2)=0$ and 
$\Phi ^{\pm \prime \prime }(0)=w_{0}$. The lower limit function $z$ is given
by 
\begin{equation}
z(x)=\theta \left( \Phi ^{-}(\alpha ,w_{0};x+\widetilde{x})+\frac{\alpha }{2}
\widetilde{w}_{0}\right) \,,  \label{zz}
\end{equation}
where $\theta <1$, $\widetilde{w}_{0}\geq w_{0}$, $\widetilde{x}=\widetilde{x
}(\alpha ,w_{0},\widetilde{w}_{0})$ is such that $z(0)=0$, with $w_{0}$ and $
\widetilde{w}_{0}$ so that $T$ is very large and $z^{\prime }(\pi )=0$ which
can always be done in view of the properties of $\Phi ^{-}$. The upper limit
function $Z$ can be written also as (\ref{zz}) with $\Phi ^{-}$ replaced by $
\Phi ^{+}$ and the second term with minus sign. We have 
\begin{equation*}
M[W^{\pm }]=\alpha \theta \{(1-\theta )(\Phi ^{\pm \prime })^{2}\mp (
\widetilde{w}_{0}-w_{0})\}\,,
\end{equation*}
with $W^{+}=Z$ and $W^{-}=z$. In order inequality (\ref{zuZ}) holds
uniformly in $\tau $, $\widetilde{u}_{xx}(t,0)$ has to remain bounded from
above and below. Since $\widetilde{u}_{xx}(t,0)<\alpha ^{-1}$ by Theorem
\ref{cone}, one may choose $\theta $ arbitrarily small in (\ref{zz}) and take
$\widetilde{w}_{0}$ and $w_{0}$ so large that $\theta (\widetilde{w}
_{0}-w_{0})>\alpha ^{-1}$. In the limit as $\theta \rightarrow 0$ we have $
\widetilde{u}(t,x)\geq 0$ and $0\leq \widetilde{u}_{xx}(t,0)<\alpha ^{-1}$
for $0\leq t\leq \tau $, uniformly in $\tau $, implying $\widetilde{u}
(t,x)\geq 0$ for all $t\geq 0$.
\end{remark}

LaSalle's invariance principle allows us to apply Liapunov function
techniques under milder assumptions. A subset $\mathcal{K}\subset \mathcal{V}
$ of a complete metric space $\mathcal{V}$ is said to be \emph{invariant}
(positive invariant) if, for any $v_{0}\in \mathcal{K}$, there exist a
continuous curve $v:\mathbb{R}\longrightarrow \mathcal{K}$ with $v(0)=v_{0}$
and $S(t)v(\tau )=v(t+\tau )$ for all $t\geq 0$ and $\tau \in \mathbb{R}$ ($
\mathbb{R}_{+}$). The following two theorems express the content of this
principle.

\begin{theorem}
\label{limitset} Suppose $v_{0}\in \mathcal{V}$ is such that the orbit $
\gamma (v_{0})=\left\{ S(t)v_{0},t\geq 0\right\} $ through $v_{0}$ lies in a
compact set in $\mathcal{V}$ and let $\omega (v_{0})$ denote its $\omega $
--limit set, i.e., 
\begin{equation*}
\omega (v_{0})=\bigcap_{\tau \geq 0}\overline{\gamma \left( S(\tau
)v_{0}\right) }\,
\end{equation*}
(see (\ref{omega}) for alternative definition). Then $\omega (v_{0})$ is
nonempty, compact, invariant, connected and $\mathrm{dist}\left( S(\tau
)u_{0},\omega (v_{0})\right) \longrightarrow 0$ as $t\rightarrow \infty $.
\end{theorem}

\noindent \textbf{Proof.} We refer to Theorem $4.3.3$ of \cite{H} for
details. Note that $\omega (v_{0})$ is the intersection of a decreasing
collection of nonempty compact sets. Note, in addition, that $\omega (v_{0})$
is positive invariant by definition and is invariant by compactness argument.

\hfill $\Box $

\begin{theorem}
\label{invariant} Let $V$ be a Liapunov function for $t\geq 0$ and, for 
\begin{equation}
\mathcal{E}:=\left\{ v\in \mathcal{V}:\,\overset{\cdot }{V}(v)=0\right\} ,
\label{EV}
\end{equation}
let $\mathcal{K}$ be the maximal invariant set in $\mathcal{E}$. \ If the
orbit $\gamma (v_{0})$ lies in a compact set in $\mathcal{V}$, then $%
S(t)v_{0}\longrightarrow \mathcal{K}$ as $\ t\rightarrow \infty $.
\end{theorem}

\noindent \textbf{Proof.} \ By definition, $V\left( S(t)v_{0}\right) $ is a
nonincreasing function of $t$ and bounded from below, by hypothesis. So, $
\lim\limits_{t\rightarrow \infty }V\left( S(t)v_{0}\right) =\upsilon $
exists. If $y\in \omega (v_{0})$, then $V(y)=\upsilon $ and, in view of the
fact that $S(t)y=y$, we have $V\left( S(t)y\right) =\upsilon $ which implies 
$\overset{\cdot }{V}(t)=0$ and $\omega (v_{0})\in \mathcal{K}$.

\hfill $\Box $

Now, we apply the invariance principle to the problem at our hand. As we
will see, if $\mathcal{B}_{0}$ is a sufficient large ball around $\phi _{0}=0
$ in the cone $\mathcal{C}$ (with the induced topology of $H_{\mathrm{e,p}
}^{1}$), the invariant set $\mathcal{K}_{k}=\left\{ \omega
(u_{0}),\,u_{0}\in \mathcal{B}_{0}\right\} \subset \mathcal{E}$ consists of
the union of unstable manifolds for the equilibrium points $\phi _{0},\phi
_{1},\ldots ,\phi _{k}$, with $\phi _{j}(x)=\displaystyle\int_{0}^{x}\psi
_{j}^{+}(y)\,dy$, provided $\alpha $ is such that $2/\left( k+1\right)
^{2}\leq \alpha <2/k^{2}$ holds for some $k\in \mathbb{N}$. Note that the
hypotheses of Theorems \ref{limitset} and \ref{invariant} hold since the
orbits of $S(t)v_{0}$ are bounded in $H_{\mathrm{o,p}}^{1}$ by Theorem
\ref{cone} and remain in a compact set of $H_{\mathrm{o,p}}^{1}$ in view of 
Theorem \ref{thivp}. For this the Sobolev embedding theorem is evoked: $
W^{2,2}\left( -\pi ,\pi \right) \subset C^{1+a}\left( -\pi ,\pi \right) $
with continuous inclusion, so $v$ has a continuous representative in $C_{
\mathrm{o,p}}^{1+a}$ which belongs to $C_{\mathrm{o,p}}^{2+a}$ by Schauder
estimates (see e.g \cite{S} and references therein). Therefore, any
solutions $\widetilde{u}(t,x)=\int_{0}^{x}v(t,y)\,dy$ of (\ref{utildeq}) in $
\mathcal{C}$ has a continuously three--times differentiable representative.

We thus have

\begin{theorem}
\label{gstb} If $\alpha >2$, $\phi _{0}=0$ is globally asymptotically stable
solution of (\ref{utildeq}) in 
\begin{equation*}
\widetilde{\mathcal{C}}=\left\{ u\in H_{\mathrm{e,p}}^{1}\cap H_{\mathrm{e,p}
}^{2}:u(0)=0,\,u\geq 0\,\,\mathrm{and}\,\,\alpha u_{xx}<1\right\} \,.
\end{equation*}
If $\alpha <2$, the origin is unstable in $\widetilde{\mathcal{C}}$ and
there exits an open dense set $\mathcal{U}\subset \widetilde{\mathcal{C}}$
of initial conditions such that $\lim\limits_{t\rightarrow \infty }
\widetilde{u}(t;u_{0})\longrightarrow \phi _{1}^{+}$ for all $u_{0}\in 
\mathcal{U}$.
\end{theorem}

\noindent \textbf{Proof.} It follows from Theorem \ref{invariant} $v(t;\cdot
)\longrightarrow \omega \left( v(0;\cdot )\right) \subset \left\{ \psi : 
\overset{\cdot }{V}\left( \psi \right) =0\right\} $ in $\mathcal{V}$ as $
t\rightarrow \infty $. But, from (\ref{Vv}), $\overset{\cdot }{V}\left( \psi
\right) =0$ iff 
\begin{equation}
\alpha \psi ^{\prime \prime }-2\alpha \,\psi \psi ^{\prime }+2\psi =0\,,
\label{phixx}
\end{equation}
whose solutions are $\psi =\psi _{0}$ and $\psi _{j}^{+},\,j=1,\ldots ,k$,
studied in Section \ref{SS}. \ We note that $\phi _{j}^{+}(x)=\displaystyle
\int_{0}^{x}\psi _{j}^{+}(y)\,dy\geq 0$ ($\phi _{j}^{-}(x)\leq 0$) for all $
x\in \lbrack -\pi ,\pi ]$ and $j\geq 1$, since $\left( \psi _{j}^{+}\right)
^{\prime }(0)>0$ ($\left( \psi _{j}^{-}\right) ^{\prime }(0)<0$).

Multiplying (\ref{phixx}) by $\psi $ and integrating over $\left( -\pi ,\pi
\right) $, gives 
\begin{equation*}
\int_{-\pi }^{\pi }\left( \alpha \psi ^{\prime \prime }+2\psi \right) \psi
\,dx=-\left\| \psi \right\| _{1/2}^{2}\leq 0
\end{equation*}
if $\alpha >2$. The nonlinear term vanishes since, by integration by parts, 
\begin{equation*}
\int_{-\pi }^{\pi }\psi ^{\prime }\psi ^{2}\,dx=-2\int_{-\pi }^{\pi }\psi
^{\prime }\psi ^{2}\,dx\,.
\end{equation*}

This implies $\psi \equiv 0$ and proves that $S(t)v_{0}\longrightarrow 0$ as 
$t\rightarrow \infty $ in $\mathcal{V}$. We quote Theorem $4.3.5$ in \cite{H}
for the instability assertion.

Since the spectrum $\sigma (L)$ of the linearized operator around the
equilibrium points (see Theorem \ref{positive}) lies on the real line, all
equilibrium points are hyperbolic, $\mathcal{E}$ given in (\ref{EV}) is a
discrete and finite set and 
\begin{equation*}
\mathcal{V=}\bigcup_{\psi \in \mathcal{E}}\mathcal{W}^{s}(\psi )
\end{equation*}
holds with $\mathcal{W}^{s}(\psi )=\left\{ u_{0}\in \mathcal{V}:S(t)v_{0}
\longrightarrow \psi \;\mathrm{as}\;t\rightarrow \infty \right\} $. It is
proven in \cite{H} that each stable manifold $\mathcal{W}^{s}(\psi )$ is a $
C^{2}$ embedded submanifold of $\mathcal{V}$ ($\mathcal{W}^{s}(\phi )$ is $
C^{3}$ submanifold of $\widetilde{\mathcal{C}}$) and, if $\psi $ is locally
unstable, than $\mathcal{W}^{s}(\psi )$ has codimension larger than or equal
to $1$. Therefore, $\mathcal{V}$, and consequently $\widetilde{\mathcal{C}}$
, can be written as a \emph{finite} union of open connected sets together
with a closed nowhere--dense remainder.

\hfill $\Box $

Finally, we show that, for an open set $\mathcal{V}_{0}\subset \mathcal{V}$
given as before, \ the maximal invariant set 
\begin{equation}
\mathcal{K}_{k}=\bigcup_{\psi \in \mathcal{E}}\mathcal{W}^{u}(\psi )
\label{Mk}
\end{equation}
where $\mathcal{W}^{u}(\psi )=\left\{ v_{0}\in \mathcal{V}%
:S(t)v_{0}\longrightarrow \psi \;\mathrm{as}\;t\rightarrow -\infty \right\} $
is the unstable manifold of $\psi $. By Theorem \ref{invariant}, the orbit $
v(t;u_{0})=S(t)v_{0}$ exists and remains, by invariance, in $\mathcal{K}_{k}$
for all $t\in \mathbb{R}$. Therefore, $\lim\limits_{t\rightarrow \infty
}v(t,u_{0})=\psi $ exists and $\psi \in \mathcal{K}_{k}$, so $\mathcal{K}
_{k}\subset \bigcup_{\psi \in \mathcal{E}}\mathcal{W}^{u}(\psi )$. Since the
converse is also true, the equality (\ref{Mk}) thus holds.

\medskip 

\begin{center}
\textbf{Acknowledgments}
\end{center}

We wish to thank J. Fernando Perez for posing to one of the authors
(D.H.U.M.), together with C. Ragazzo, the problem of what is the effect of
the sequence of thresholds on the stable branch. The authors have benefited
from the discussions with C. Ragazzo, W. F. Wreszinski and J. C. A. Barata.

\end{document}